# Nanostructured Polymer Films with Metal-like Thermal Conductivity

*Yanfei Xu[1], Daniel Kraemer[1], Bai Song[1], Zhang Jiang[2], Jiawei Zhou[1], James Loomis[1], Jianjian Wang[1], Mingda Li[1], Hadi Ghasemi[1†], Xiaopeng Huang[1], Xiaobo Li[1†], Gang Chen[1]\**

[1]Department of Mechanical Engineering, Massachusetts Institute of Technology, Cambridge, Massachusetts 02139, USA.

[2]Advanced Photon Source, Argonne National Laboratory, Argonne, Illinois 60439, USA.

*Correspondence to: gchen2@mit.edu

†Current address:

Department of Mechanical Engineering, University of Houston, Houston, Texas 77004, USA.

State Key Laboratory of Coal Combustion, School of Energy and Power Engineering, Huazhong University of Science and Technology, Wuhan, Hubei 430074, P. R. China.

The table of contents entry: Scalable transformation of plastics into excellent heat conductors enabled by insight into transport mechanisms associated with amorphous polymer chains.

Abstract:

Thermally conductive polymers are of fundamental interest and can also be exploited in thermal management applications. Recent studies have shown stretched polymers can achieve high thermal conductivity. However, the transport mechanisms of heat in thermally conductive polymers have yet to be elucidated. Here we report a method for scalable fabrication of polyethylene films with a high thermal conductivity of 62 W/m-K. The achieved thermal conductivity is over two orders-of-magnitude greater than that of typical polymers (~0.1 W/m-K), and exceeds those of many metals and ceramics used as traditional heat conductors. Careful structural studies are carried out and reveal that the film consists of nanofibers with crystalline and amorphous regions. Contrary to conventional wisdom, we reveal the importance of the amorphous morphology in achieving such high thermal conductivity, rather than simply from enhancements in the degree of crystallinity and crystallite alignment. The amorphous phase reaches a remarkably high thermal conductivity of ~16 W/m-K. Even still, we identify that the presence of this amorphous phase is the dominant factor as the film thermal conductivity is still much lower than the predicted values for bulk single-crystal polyethylene (237 K/m-K). This work lays the foundation for the rational design and synthesis of thermally conductive polymers, and opens up new opportunities for advanced heat management, particularly when flexible, lightweight, chemically inert and electrically insulating thermal conductors are desired.

Main Text:

From soft robotics, organic electronics to 3D printing and artificial skin, polymers continue to infiltrate modern technologies thanks to their unique combination of properties not available from any other known materials.[1–5] They are lightweight, durable, flexible, corrosion resistant, and easy to process, and hence are expected to offer significant advantages over traditional heat conductors such as metals and ceramics.[1] However, application of polymers in thermal management has been largely hampered by their low thermal conductivities (~0.1 W/m-K).[6] To date, metals and ceramics remain the dominant heat conductors.

The fact that polyethylene (0.2-0.5 W/m-K)[6,7] is composed of a backbone of carbon-carbon bonds similar to those in diamond, one of the most thermally conductive materials (above 1000 W/m-K),[8] encourages research in thermally conductive polymers. Importantly, atomistic simulations have suggested that an individual crystalline polyethylene chain can achieve very high — possibly divergent — thermal conductivity,[9] in agreement with the non-ergodic characteristics of one-dimensional conductors discussed by Fermi, Pasta and Ulam.[10] However, the experimental measurement of such theoretically high thermal conductivities remains elusive. By increasing the crystallite orientation and crystallinity, the thermal conductivity of polymers can increase considerably,[11–17] such as polyethylene nanofibers (~104 W/m-K) and polyethylene fibers (~20 W/m-K).[15] Although exceptionally conductive, these measured values are still much lower than the numerical predictions for bulk single-crystalline polyethelyene (~237 W/m-K).[18] There is no precise mechanism that accounts for the deviation of experimental and theoretical values. And the main factors that govern the thermal conductivity in these fibers remain poorly understood.[19] It is generally known that such materials are not perfect crystals, but instead semicrystalline polymers containing mixed crystalline and amorphous regions.[6] Translating the remarkably high thermal conductivity seen in simulation as well as in polyethylene nanofibers into a scalable polymer presents a major challenge in synthesis. Overcoming this challenge will broadly expand the scope of nanofiber use in thermal management, since practical applications require large areas or volumes of materials.[20] Recently, Ronca et al. reported stretched ultrahigh molecular weight film with thermal conductivity as high as 65 W/m-K,[16] measured using a commercial laser-flash system, and Zhu et al. reported thermal conductivity of fibers as high as 51 W/m-K by further processing of commercial Spectra fibers using an electrothermal method.[17] These reports show the potential of achieving high thermal conductivity in macroscopic samples. However, the structural property relationship has yet to be further elucidated.

We have been engaged in scaling up the high thermal conductivity of individual nanofiber to more macroscale films.[21] Here, we report a thermal conductivity measurement of 62 W/m-K in polyethylene films (Fig. 1). The thermal conductivity in our film outperforms that of many conventional metals (304-stainless steel ~15 W/m-K) and ceramics (aluminum oxide ~30 W/m-K). Motivated by the theoretically large thermal conductivity of single-crystal polymer,[9,18] we fabricated thermally conductive polymer films with an emphasis on minimally entangling and maximally aligning the chain, rather than solely pursuing a high crystallinity. We further uncover the thermal transport mechanisms through the combination of structural analysis, determined by high-resolution synchrotron x-ray scattering, and a phenomenological thermal transport model. We find that the film actually consists of nanofibers with crystalline and amorphous regions along the fiber and that the amorphous regions have remarkably high conductivity (~16 W/m-K), which is central to the high thermal conductivity (62 W/m-K). Increased control over amorphous morphology is a promising route towards achieving thermal conductivities approaching theoretical limits.

We start with commercial semi-crystalline polyethylene powders (**Fig. 1**a) which feature randomly oriented lamellar crystallites (lamellae) dispersed in an amorphous chain network (Fig. 1d). We dissolved the powder above its melting temperature in decalin, allowing the initially entangled chains to disentangle (Fig. 1d). This greatly reduced the entanglements for the subsequent processing. Afterwards, the hot solution was extruded through a custom-built

Couette-flow system,[21] which imparted a shear force on the polymer chains and led to further disentanglement.[22] To maintain the disentangled structure, the extruded solution flows directly onto a liquid nitrogen-cooled substrate. Some segments of the polyethylene chains folded back into thin lamellae upon drying,[7] while others remained disordered albeit less entangled (Fig. 1d).[23] Finally, the as-extruded films (Fig. 1b) were mechanically pressed and drawn inside a heated enclosure using a continuous and scalable roll-to-roll system.[21] Heating allowed the disentangled polymer chains to move more freely and facilitated alignment along the draw direction(Fig. 1c, d).[6]

In order to track the evolution of polymer structures, we imaged the as-purchased powders, the extruded films and films of various draw ratios (final length / initial length) using scanning electron microscopy (SEM, Fig. 1e-j). The powder consists of porous particles with an average size of ~100 $\mu$m (Fig. 1e). After extrusion, the film surface appeared isotropic with randomly distributed microflakes (Fig. 1f). During drawing, the film self-organized into a clear fibrous texture along the draw direction. The diameters of the fibers comprising the film reduced as the draw ratio increased, which led to a smoother and denser texture (see 10× and 110× in Fig. 1g, h). We further tore a 70× film apart to explore the detailed internal structures where individual fiber can be clearly observed (Fig. 1i, j), and multiple interior fibers with smaller diameter ~ 8nm were also seen (Supplementary Information, SI, SEM).

To study the thermal properties of these polyethylene films, we employed two distinct experimental schemes: a home-built steady-state system[24] (**Fig. 2**a) and a widely-adopted transient method called time domain thermoreflectance (Fig. 2c, SI thermal conductivity measurements).[25–27] On the steady-state platform, we measured heat current as a function of temperature difference across a sample film. To validate the steady-state measurement accuracy, we first tested several control samples including 304-stainless steel,[28] Zylon and Dyneema fibers (Fig. 2b),[15] and obtained 15.3 W/m-K, 22.6 W/m-K and 23.6 W/m-K, respectively, which were in excellent agreement with established values.[15,28] We next measured the thermal conductivities of a series of films with various draw ratios (**Fig. 3**a). The as-extruded (1×) film was found to have an in-plane thermal conductivity of 0.38 W/m-K. As draw ratio was increased, film thermal conductivity along the draw direction was improved dramatically, reaching 62 W/m-K at 110× (Fig. 2b). Notably, we saw no sign of saturation in thermal conductivity (Fig. 3a and Fig. 3b will be discussed below), which suggested more room for further conductivity enhancement beyond 110× draw ratio. Recent atomistic simulations further corroborate this expectation.[18]

Two-color time-domain thermoreflectance (TDTR) experiments were conducted to study transient heat conduction in the films,[25–27] and to further validate the steady-state results (Figs. 2c). We fabricated a 150 μm-thick laminate consisting of 100 layers of 50× films and carefully microtomed a cross-section (roughness ~10 nm, Fig. S6) perpendicular to the draw direction. Representative thermoreflectance signals are reported in Fig. 2d, from which we extracted an average thermal conductivity of 33.6 W/m-K along the draw direction. The TDTR results agree well with values obtained using the steady-state system (Fig. 3a). The successful demonstration of 100-layer laminate with such high thermal conductivity implies potential scalability not only along the drawing direction, but also in the thickness direction. In addition, we have investigated film thermal stability, obtaining <5% thermal conductivity variations before and after annealing (24 hours at 80 °C).

To reveal correlation between such a high thermal conductivity and structure, we quantitatively investigated the structure at both atomic scale and nanoscale by high-resolution wide-angle and small-angle synchrotron x-ray scattering (**Fig. 4** and SI structure characterization). Wide-angle x-ray scattering (WAXS) measurements were used to determine the crystallite orientation and crystallinity. Comparisons between the as-extruded and drawn films show a clear transition from concentric rings characteristic of polycrystalline samples to short arcs (10×), which eventually become discrete spots (110×), suggesting improved alignment of initially randomly-oriented crystallites upon drawing (Fig. 4b). Specifically, the initially isotropic peaks in the {*hk0*} group became narrow and oriented along the meridian direction c, indicating that the *c*-axis (chain direction, Fig. 4a) aligns with the draw direction. The degree of orientation was quantified via an intensity-weighted average over the angle between the *c*-axis and the draw direction (Fig. 4a).[29] The orientation order parameter quickly increases from zero for as-extruded films to nearly saturated value for perfectly aligned crystals at a draw ratio as low as 2.5× (Fig. 4d). The thermal conductivity of the 2.5× films (4.5 W/m-K) was over 10 times larger than the as-extruded ones (0.38 W/m-K, Fig. 3). We therefore expect the excellent alignment of the crystallites to be responsible for the limited thermal conductivity enhancement at very low draw ratio, which is consistent with the conventional strategies to improve the thermal transport in polymers.[11]

However, after 10× draw ratio where the orientation factor nearly saturates, we observed an additional 10-fold thermal conductivity enhancement to the 62 W/m-K (110×), which clearly suggested other enhancement mechanism. We noticed that during the stretching the crystallinity first increased at a high rate at low draw ratios (below 10×) and then steadily grew to over 90% in 110× films (Fig. 4d). Contrary to the past work that emphasized on crystallinity-dependence of the thermal conductivity,[12] the weak growth rate of crystallinity at high draw ratios is clearly not sufficient to account for the dramatic boost of the thermal conductivity, and there is even no sign of saturation of the conductivity as the draw ratio increases (Fig. 3).

These observations convinced us that unknown factors other than the crystalline phase play the crucial roles especially at high draw ratios. We therefore resorted to the structures of the amorphous region for clues. Quantitative analysis of small-angle X-ray scattering (SAXS) intensity profiles along the draw direction reveals two humps at scattering vectors that differ by a factor of two (Fig. 4d), indicating a periodic structure with a repeating unit consisting of alternating crystalline and amorphous phases (Figs. 1d and Fig. S10).[30] This picture agrees with the widely-accepted lamella-like structural model for stretched polyethylene.[7] The displacement of the humps towards smaller scattering vectors with increasing draw ratio indicates that the period length grows with drawing. Normalized electron density profiles further reveal the relative lengths of crystalline and amorphous regions in each unit (Fig. 4f inset). Specifically, the amorphous fraction decreases with increasing draw ratio (Fig. 4f), consistent with the increasing trend of crystallinity (Fig. 4d) and film thermal conductivity (Fig. 3a). However, decreasing the fraction of amorphous region alone cannot account for such observed ultrahigh conductivity, because amorphous phase is simply too thermally resistive.[6]

To provide further evidence of the dominant role of amorphous region, a phenomenological one-dimensional thermal transport model is developed (SI thermal model). Based on the structural parameters obtained in WAXS and SAXS, the crystalline and amorphous regions are randomly mixed in the as-extruded films. Upon stretching, aligned fibers consisting of alternating crystalline and amorphous regions are developed in the interior of the film. The average fiber diameter was estimated to be ~10-50 nm nanometers (Fig. 1i, j, Fig. S3c and Fig. S11c), justifying the use of a one-dimensional model $k = [(1-\eta)/k_c + \eta/k_a]^{-1}$ for the axial thermal conductivity. Here $\eta$ is the amorphous fraction in a periodic unit, and can be fitted from the SAXS analysis (Fig. 4f), while $k_c$ and $k_a$ are the thermal conductivities of the crystalline and amorphous regions, respectively. The crystalline thermal conductivity $k_c$ depends on the crystallite size due to phonon scatterings at boundaries, and is estimated based on a recent first-principles calculation for a one-dimensional polyethylene chain[18]. Combined with our measured total thermal conductivity, these yield the amorphous thermal conductivity $k_a$ as the draw ratio (Fig. 3b). It is clearly seen that the experimentally measured high thermal conductivities at higher draw ratios suggest a high $k_a$ (~5 W/m-K at 50×, and 16 W/m-K at 110× versus typical 0.3 W/m-K). In other words, the amorphous region after drawing is no longer composed of random disordered chains, but rather has developed some degrees of the orientation order with more extended and aligned chains. This is also consistent with our experimental observation that the isotropic amorphous diffusing ring gradually disappeared from 10× to 110× (WAXS, Fig. 4b and Fig. S12), and consistent with the Raman study by Zhu et al. on further stretched Spectra fiber.[17] The extracted high thermal conductivity of the amorphous region with some molecular orientation is much higher than that of oriented polythiophene fibers grown in a template,[19] despite theoretical prediction of higher thermal conductivity of polythiophene than polyethylene in crystal form.[31] Thus, thermal resistance of amorphous region ultimately governs the overall thermal conductivity. With this in mind, our findings here will steer the research to a new track — engineering amorphous chains to achieve even higher thermal conductivity.

In summary, conventional efforts that focus on crystalline phase in polymers can only marginally increase the thermal conductivity. In contrast, we have elucidated that improvement of the amorphous phase is the critical step for heat transfer enhancement. The amorphous region in our film has remarkably high thermal conductivity, but remains the dominant resistance, and it is where further research effort would be most effectively focused. We have developed a scalable mass production route for producing polymer films with metal-like thermal conductivity. The past few years have witnessed a surge in the interest of using polymers for thermal management and energy conversion. We believe that the high thermal conductivity achieved in these polymer films, with their unique combination of characteristics (light weight, optical transparency, chemical stability etc.) will play a key role in many existing and unforeseen applications. Of course, polyethylene itself has limitations in the temperature range it can cover. We foresee that further improvement of the thermal conductivity of the persistent amorphous phase will be the key developing the next generation of heat-conducting polymers, in polyethylene and beyond.

**Acknowledgements**

The authors acknowledge support from Department of Energy/Office of Energy Efficiency & Renewable Energy/Advanced Manufacturing Program (DOE/EEREAMO) under award number DE-EE0005756 (prior to 2/2016); the MIT Deshpande Center, and ONR MURI under award number N00014-16-1-2436. This research used resources of the Advanced Photon Source, a U.S. Department of Energy (DOE) Office of Science User Facility operated for the DOE Office of Science by Argonne National Laboratory under Contract No. DE-AC02-06CH11357. The authors thank T. Sanchez and N. Liu from MIT undergraduate research opportunities program for sample preparation; G. Ni for the photograph help; W. Dinatale and L. Wu for SEM discussions; C. Marks and D. Bell for microtome discussions at the Center for Nanoscale Systems, Harvard University; N. Thoppey for his participation in the DOE project; S. Huberman, L. Meroueh and V. Chiloyan for time-domain thermoreflectance discussions; C. Settens (CMSE, MIT), H. Li (Northeastern University), M. Minus (Northeastern University), S. Billinge (Columbia University) and M. Terban (Columbia University) for providing X-ray diffraction testing and discussions; G. Ni, J. Tong, S. Boriskina, T. Cooper, Y. Huang, Q. Song, and L. Weinstein for emittance discussions. S. Huberman, V. Chiloyan, J. Mendoza, L. Meroueh, G. Ni, T. Cooper for the discussion and proofreading.

Author Contributions: The materials were fabricated by Y. Xu. The material thermal properties were characterized by D. Kraemer, B. Song, Y. Xu and J. Zhou. The material structures were characterized by Y. Xu and Z. Jiang. Structural and thermal modelling were performed by Z. Jiang and J. Zhou, respectively. Custom-built continuous production platform was built by J. Loomis. J. Wang, M. Li, H. Ghasemi, X. Huang, X. Li participated in different phases of this project and contributed to the discussion and understanding of the materials. The manuscript was written by Y. Xu, B. Song and Z. Jiang with comments and inputs from all authors. G. Chen directed the research.

Additional Information: patents have been filed: continuous fabrication platform for highly aligned polymer films, WO2015171554 A1.

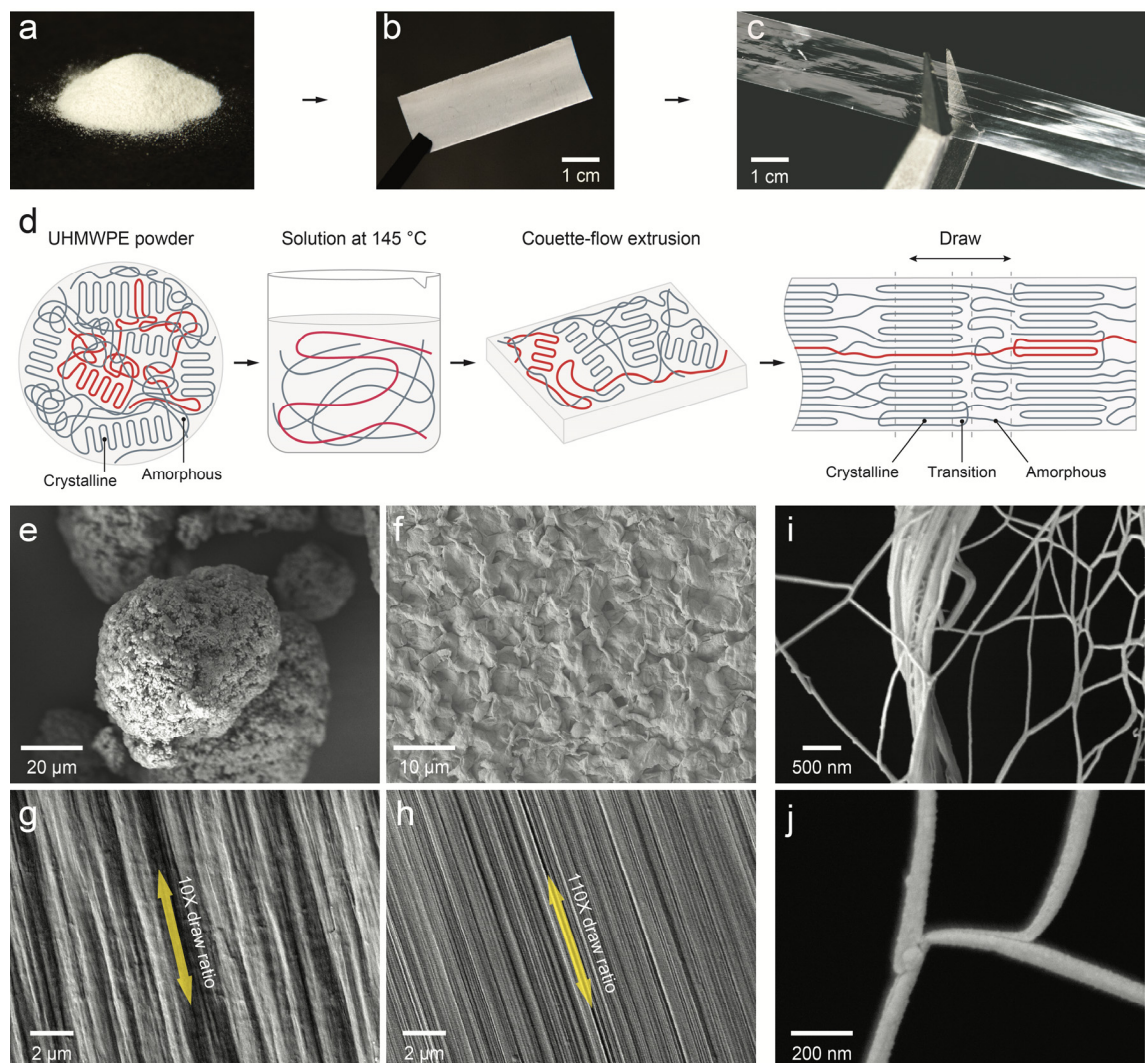

**Fig. 1. Fabrication and characterization of novel polymer films with high thermal conductivity.** (**a-c**) Photos of commercial ultrahigh molecular weight polyethylene (UHMWPE) powders, a thick opaque as-extruded film and a thin transparent drawn film, respectively. (**d**) Illustration of film morphology evolution during fabrication. The powders feature lamellar polyethylene crystallites embedded in a disordered and entangled chain network. The degree of entanglement greatly reduces in the hot decalin solution and after subsequent Couette-flow extrusion. The ultradrawn films are characterized by oriented crystallites interconnected by aligned amorphous chains. (**e-h**) Scanning electron microscope (SEM) images of some UHMWPE powders, an as-extruded film, a 10× draw ratio film and a 110× film. (**i-j**) SEM images of a torn 70× film revealing the polyethylene nanofibers as the basic building blocks.

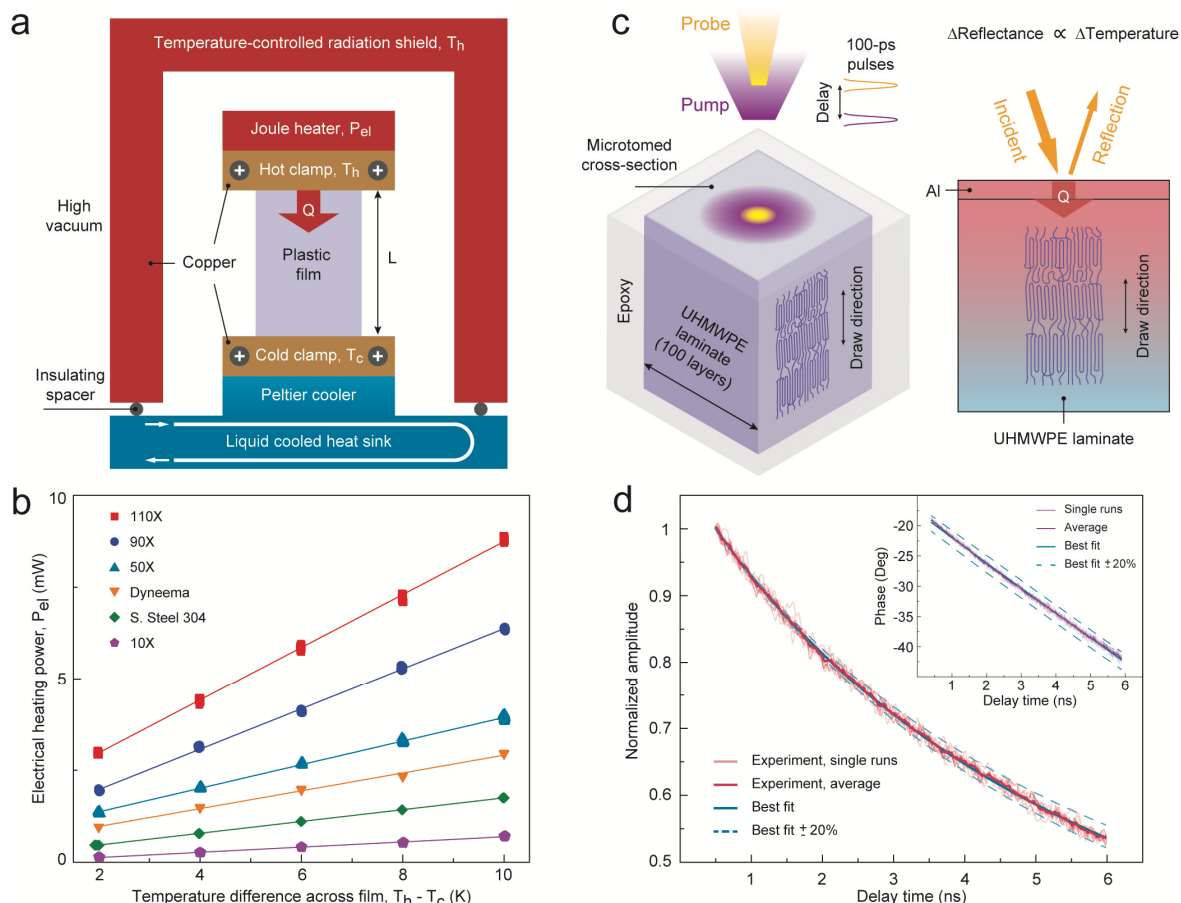

**Fig. 2. Measurement of heat transport along the draw direction of the polymer films.** (**a**) Schematic of the home-built steady-state thermal conductivity measurement system. A small temperature difference ($T_h$-$T_c$) across a film sample is created and maintained using Joule heating ($P_{el}$) and thermoelectric cooling inside a high vacuum chamber. (**b**) The electrical heating power $P_{el}$ applied to maintain a constant $T_h$ while $T_c$ is varied. Representative data as scaled to the geometry of a 50× film, $P_{el}$ was measured multiple times at each $T_h$-$T_c$. A larger slope indicates a higher thermal conductivity (SI). (**c**) Illustration of the two-color time-domain thermoreflectance measurement scheme. An aluminum-coated UHMWPE laminate is first heated with a 100-fs-wide pump laser pulse (400 nm, purple) and subsequently monitored with a time-delayed low-power probe pulse (800 nm, yellow). The change in aluminum reflectance is proportional to surface temperature variation in the linear regime. (**d**) Ten individual cooling curves in terms of signal amplitude (light red lines), overlaid with their average (thick red) and the best fit curve (blue solid) that yields a thermal conductivity of 31.9 W/m-K (SI). Changing the best fit by 20% leads to large discrepancies between the simulated (blue dashed) and measured curves. Inset shows the corresponding phase signals, fitting to which yields a thermal conductivity of 32.8 W/m-K.

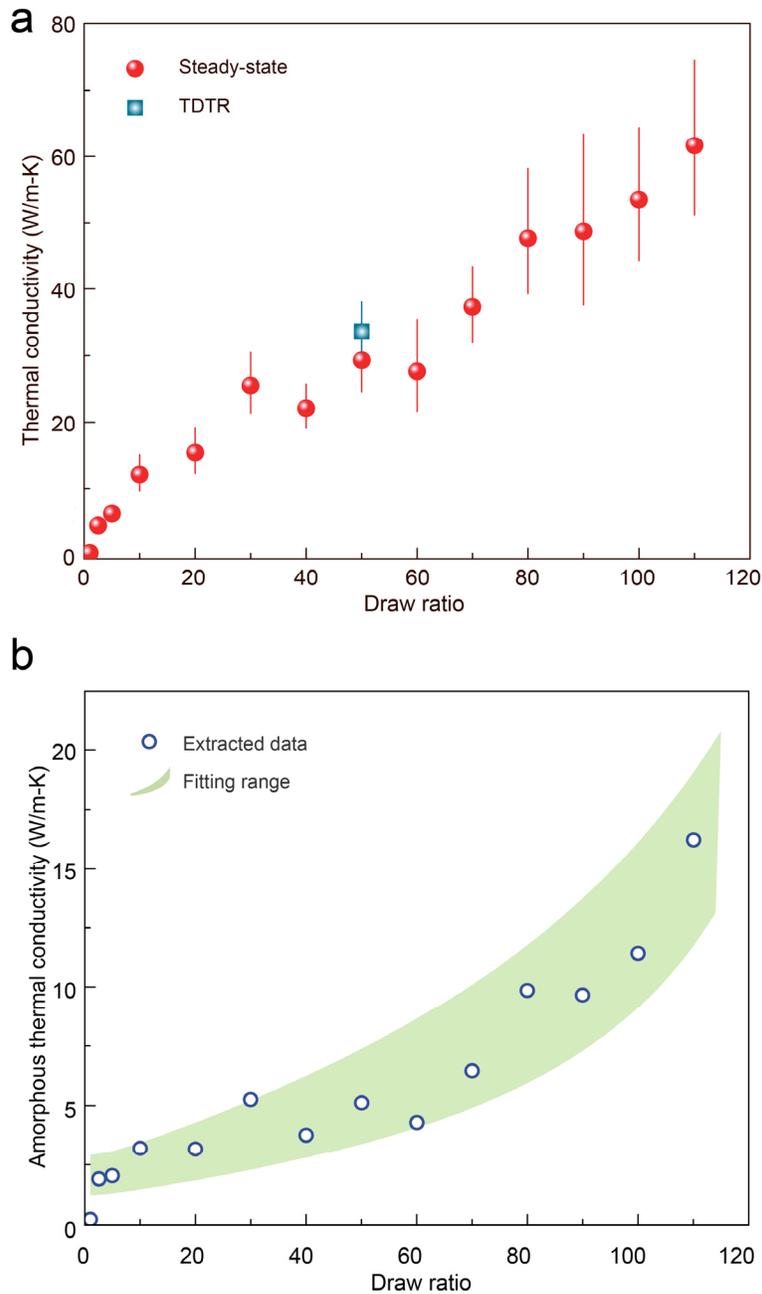

**Fig. 3. Measured and computed thermal conductivities for the polymer films**. (a) Measured total thermal conductivity as a function of draw ratio. The red spheres were obtained from the steady-state experiments. A thermal conductivity of 62 W/m-K was measured from the 110× films. The blue square denotes the average of 20 transient thermoreflectance measurements at 3 MHz and 6 MHz modulation. (b) Extracted amorphous thermal conductivity values based on fitted structural parameters from SAXS analysis. The dots are calculated using the measured total thermal conductivities. The shaded region is obtained by fitting the total thermal conductivity with a straight line, and further adding the uncertainties in the determination of structural parameters, thereby giving the estimation of the amorphous thermal conductivity between the upper and lower bound (see more details in SI).

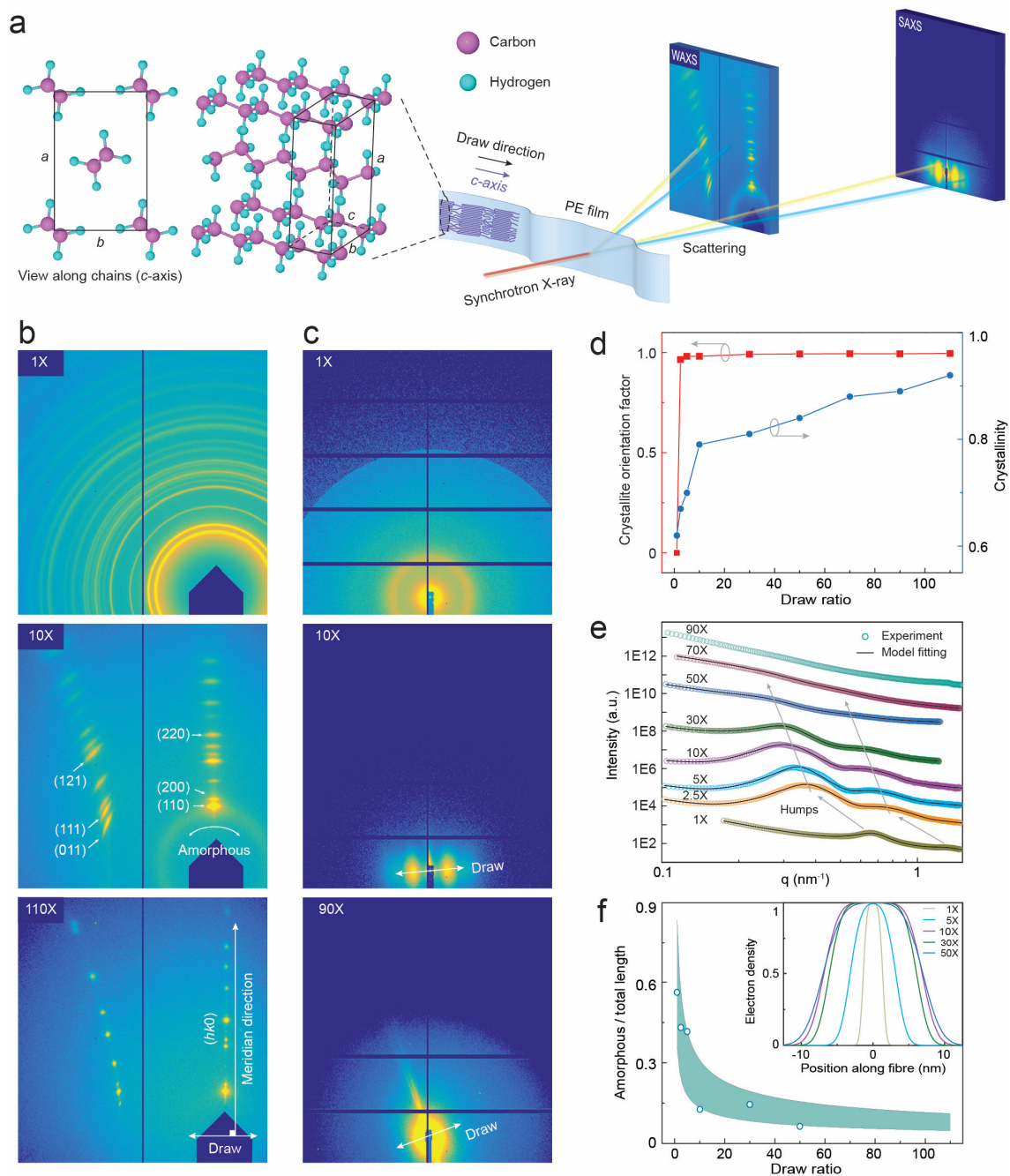

**Fig. 4. Structural characterization using synchrotron X-ray scattering.** (**a**) Illustration of the experimental setup and the orthorhombic unit cell of crystalline polyethylene. The incident beam is perpendicular to the drawn direction. The lattice constants were obtained as a = 7.42, b = 4.95, c = 2.54 Å, where c-axis is the chain direction. (**b**) Wide angle X-ray scattering (WAXS) patterns from the 1×, 10× and 110× films. Characteristic Bragg scattering by the {hk0} and {hk1} plane groups were observed. The {hk0} group appears perpendicular to the draw direction. (**c**) Small angle X-ray scattering (SAXS) patterns from the 1×, 10× and 90× films, which clearly show an isotropic-to-anisotropic transition. (**d**) First-order orientation parameter and the effective crystallinity obtained from WAXS. (**e**) Scattering intensity linecuts of the SAXS patterns along the draw direction. Two humps appeared at scattering vectors that differ by a factor of two, suggesting a periodic structure with a repeating unit consisting of alternating crystalline and amorphous phases (SI). The humps moved towards a smaller q with increasing draw ratio, indicating an increase in the period length. (**f**) The fraction of amorphous region in one periodic unit as a function of draw ratio (Fig. S14). The circles were directly extracted from the SAXS data, while the shaded zone marked the range (±40%) of fitted data which were used in the one-dimensional thermal model (SI). Inset is the normalized electron density profile obtained from SAXS analysis (SI, SAXS analysis).

# Supporting Information

**Nanostructured Polymer Films with Metal-like Thermal Conductivity**

*Yanfei Xu[1], Daniel Kraemer[1], Bai Song[1], Zhang Jiang[2], Jiawei Zhou[1], James Loomis[1], Jianjian Wang[1], Mingda Li[1], Hadi Ghasemi[1†], Xiaopeng Huang[1], Xiaobo Li[1†], Gang Chen[1\*]*

[1]Department of Mechanical Engineering, Massachusetts Institute of Technology, Cambridge, Massachusetts 02139, USA.
[2]Advanced Photon Source, Argonne National Laboratory, Argonne, Illinois 60439, USA.
\*Correspondence to: gchen2@mit.edu
†Current address:
Department of Mechanical Engineering, University of Houston, Houston, Texas 77004, USA.
State Key Laboratory of Coal Combustion, School of Energy and Power Engineering, Huazhong University of Science and Technology, Wuhan, Hubei 430074, P. R. China.

# 1. Supplementary methods

## 1.1 Fabrication of thermally conductive polyethylene films

Polyethylene solution was prepared by adding ultrahigh molecular weight polyethylene powders (UHMWPE, molecular weight 3-6 million Da, 3 wt%) into decalin solvent, and was heated to 145 °C for 24 hours in a silicone oil bath. Subsequently, the hot solution was extruded onto a liquid nitrogen-cooled glass substrate using a high-shear-rate Couette-flow extrusion system.[21] The majority of solvent was evaporated at ambient conditions for ~36 hours. The as-extruded films were then removed from the substrate and fed to a constant-force adaptive thickness roll-to-roll drawing platform (heated, ~90 °C)[21]. The draw ratios were obtained as the ratio of final to initial film length, with ~20% uncertainty.

## 1.2 Thermal conductivity measurements

Thermal conductivity measurement of the plastic films is very challenging because the samples are very thin at high draw ratio (~1-3 microns). A recent experimental study of Dyneema, Zylon and Spectra fibers suggested that some previous work may have overestimated thermal conductivities by as much as a factor of 3.[15] We explored numerous methods to measure thermal conductivity of the drawn films, and eventually decided to use an extensively validated home-built steady-state platform.[24,32,33] We calibrated the setup by measuring various control samples, including 304-stainless steel foils,[28] Dyneema fibres[15] and Zylon fibres[15], obtaining thermal conductivity values similar to those reported in previous references.[15,28] We also used a transient pump-probe technique[25–27] to measure a thick laminate and the results are consistent with the steady-state method.[24,32,33] Each method is described in more detail below.

### 1.2.1 Steady-state method
*Measurement principle*

Direct measurement of the electrical heating power ($P_{el}$) as a function of temperature difference ($T_h$-$T_c$) across a sample film was performed.[24] $T_h$ (303 K) was kept constant via feedback control of $P_{el}$, while $T_c$ was reduced to create a small temperature difference (up to 10 K) by systematically increasing the thermoelectric cooling power. Multiple measurements of $P_{el}$ were performed at a given temperature difference once the system had reached steady state (Fig. 2b). Subsequently, the slope of the linear fit yields, according to Fourier's law and after correction for thermal shunting (Fig. S1). Convection and parasitic heat loss were minimized using high vacuum and a temperature-controlled copper shield, respectively. Special effort was taken to minimize the thermal radiation exchange and to ensure that the reported thermal conductivity is conservative even if any residual radiation exists. Thermal shunting was minimized and quantified after each experiment by removing the film sample and repeating the measurement.

We create a geometry such that the one-dimensional (1D) Fourier law of heat conduction is satisfied.[24,32,33]

$$Q = \frac{kA}{L}(T_h\text{-}T_c) = \frac{kA}{L}\Delta T \quad (1)$$

where $A$, $k$, $L$, $T_h$ and $T_c$, $Q$ are the sample cross-sectional area, thermal conductivity, sample length, hot side temperature, cold side temperature and heat flow through the sample, respectively. Figure 2a and Fig. S1a show a schematic and some images of the experimental setup, respectively. The sample was suspended between a hot junction clamp and a cold junction clamp, all of which were guarded by a copper radiation shield. We maintained the hot clamp temperature $T_h$ constant using a resistive electrical heater while the cold clamp temperature $T_c$ was systematically lowered by a thermoelectric cooler. We achieved this by feed-back controlling the power input $P_{el}$ into the heater as the cooling power increased (Fig. S1b). By keeping the hot clamp and copper radiation shield at the same constant temperature, parasitic heat losses such as through the electrical leads to the heater and thermocouples were effectively kept constant, which therefore did not affect the slope of $P_{el}$ versus the temperature differential $\Delta T$. The thermal conductivity of the sample was obtained from Eq. (2).

$$k = \frac{L}{A}\frac{dP_{el}}{d(\Delta T)} \quad (2)$$

### *Analysis of uncertainties due to thermal radiation*

This method only leads to accurate results when the surrounding temperature is constant to ensure constant parasitic heat losses from the hot side and radiation loss is minimal. For these purposes, a temperature-controlled copper radiation shield guarded the heater clamp and the suspended sample. The radiation shield was constantly maintained at the hot side temperature, $T_h$, and all heater current leads and hot side thermocouple wires were thermally grounded to minimize parasitic heat losses which was especially important for samples with small thermal conductance. Radiation heat transfer between the sample and the environment is of major concern for the accuracy of the thermal conductivity measurements. Two measures were taken to ensure the reliability of reported data. As already mentioned, the sample was surrounded by a temperature-controlled (at $T_h$) radiation shield. This configuration led to an additional heat input from the shield into the suspended sample, which reduced the required electrical power input $P_{el}$ for a given temperature difference and hence the reported thermal conductivity is a more conservative value. Second, the sample length was chosen to ensure that heat conduction along the sample is higher than the surface radiation loss. The ratio of radiation to conduction heat flow can be approximated as:

$$\frac{Radiation}{Conduction} = \frac{L^2 \varepsilon \sigma ((3T_h + T_c)/2)^3}{kt} \quad (3)$$

where $t$ is the sample thickness, $k$ the thermal conductivity, $L$ the sample length, $\varepsilon$ the emittance, $\sigma$ the Stefan-Boltzmann constant, $T_h$ and $T_c$ the hot and cold side temperatures of the sample.

We attempted to measure emittance based on reflectance and transmittance values, but the low emittance and experimental uncertainties sometimes led to negative values. Therefore, we decided to use emittance of the films at different draw ratios (different thickness) based on the following empirical formula found in literature:[34]

$$\varepsilon = 2.51 \times 10^{-3} t - 3.12 \times 10^{-6} t^2 + 4.55 \times 10^{-6} tT - 4.13 \times 10^{-4} T + 0.206 \quad (4)$$

where $T$ is the film temperature in Celsius, and $t$ is the sample thickness in microns which were carefully measured using a micrometer and a profilometer, as detailed in the section below (Fig. S2a and S2b).

Figure S2c shows the films' emittance at 298 K based on Eq. (4). Figure S2d shows the right-hand side of Eq. (3) for the measured films at different draw ratios. In general, radiation errors are less than 20%, and for the higher draw ratios, less than 10%. Further, thermal radiative shunting between the hot and cold side was carefully considered. We minimized radiative shunting by using polished copper ($\varepsilon$ < 0.05) for the hot and cold clamps (Fig. S1a). In addition, radiative thermal shunting calibration was performed after each sample measurement. The same measurement procedure was used without a suspended sample allowing to obtain the radiation heat flow between the hot and the cold side. The sample geometries were optimized such that the radiative thermal shunting is limited to less than 10% relative to heat conduction by the sample.

*Analysis of uncertainties due to sample geometry*

Error in determining the drawn film thickness was minimized by using a Bruker DektakXT stylus profilometer, which was first calibrated by a 45-nm step height standard from Bruker company. The drawn film was mounted on a silicon wafer, and the film thickness was obtained by the edge step. Figure S2a shows film thicknesses at different draw ratios. Each film thickness was measured 10 times at different locations along the sample. Figure S2b shows some representative thickness profiles for the 110× film. The thicknesses of the drawn polyethylene films from 10× to 110× are ~1- 8 $\mu m$, which were measured using a profilometer. The as-extruded (1×), 2.5× and 5× film thicknesses are 76.4 $\mu m$, 16.5 $\mu m$, and 14.3 $\mu m$, respectively, as measured using a micrometer. Length and width of the film were measured after it was suspended between the clamps. Briefly, we first took a picture of the film and clamps under a microscope. The precisely machined heater clamp provides a scale bar to which the sample length is compared. This gives a reasonable accuracy of ±50 $\mu m$. Measured sample dimensions including the lengths and cross-sectional areas are listed in Table S1.

*Platform calibration using control samples*

Prior to measuring the polyethylene films, we measured control samples including 304-stainless steel foils,[28] Zylon fibres[15] and Dyneema fibres,[15] obtaining thermal conductivities of ~15.3 (+1.1, -0.6) W/m-K, ~22.6 (+5.0, -4.5) W/m-K, and ~23.6 (+4.4, -3) W/m-K, respectively, which are in good agreement with literature values.[15,28] Here, the thickness of the 304-stainless steel foils was measured by a micrometer to be 50 $\mu m$, and the cross-sectional area was 0.47 mm.[2] The Zylon sample is a bundle of 664 individual fibers. The Dyneema sample consists of 300 fibers. The average diameter of a single Dyneema fiber and Zylon fiber are 17 $\mu m$ and 11.7 $\mu m$, respectively, which were characterized with SEM (Fig. S3). The estimated radiation to conduction ratio for the measured Dyneema bundle was <2%.

*Minimization of thermal interface resistance*

Temperatures of the hot and cold sides were measured by thermocouples attached to the hot and cold clamps. Key to accurately measuring the temperatures is the use of proper thermal interface materials between the film and the copper clamps. Several thermal interface materials (TIM) were investigated using the 304-stainless steel thin foil (50 µm) as the reference sample, which has similar thermal conductance as our polymer films. We found that a certain type of unhardened silver epoxy paste provided a reliable contact and the smallest interface thermal resistance, as compared to other TIMs such as indium foil or silicone-based thermal paste. Using the unhardened silver epoxy paste also gave us the opportunity to disassemble the setup without damaging the sample and with some epoxy left on the clamped ends of the sample as marks. The distance between the marks corresponds to the length of the sample and allows us to double-check our initial length measurement (via the microscope image).

*Effect of the anisotropic thermal conductivity of the films*

One of the main characteristic properties of the drawn polymer films is the large anisotropy in thermal conductivity. In the in-plane draw direction the thermal conductivity ($k_{IP}$) is drastically larger than the perpendicular directions such as the cross-plane thermal conductivity ($k_{CP}$). This inevitably affects the temperature profile in the clamped region as shown schematically in Fig. S4a. Mean sample temperature to the left of the cold clamp can be significantly higher compared to the clamp temperature. In order to estimate the effect we approximate the two-dimensional (2D) fin conduction problem with a 1D fin problem by lumping the sample's cross-plane thermal resistance into an effective thermal contact resistance $r_{th,c} \approx t/4k_{cp}$ (Fig. S4b). The resulting differential fin equation is

$$k_{IP}\frac{t}{2}\frac{d^2T}{dx^2} - \frac{k_{CP}}{t/4}(T-T_c) = 0 \qquad (5)$$

with $t$ being the film thickness, $T$ and $T_c$ being the sample temperature along $x$-direction and the cold clamp temperature, respectively. The differential equation can be solved with a specified heat flux, $q$, at the left side of the clamp and the sample temperature being $T_c$ at infinity as the two boundary conditions.

$$T(x) = T_c + \frac{qt}{\sqrt{2k_{IP}k_{CP}}}e^{-\sqrt{\frac{k_{CP}}{k_{IP}}\cdot\frac{8}{t^2}}x} \qquad (6)$$

The introduced error in the thermal conductivity measurement stems from the difference in the copper clamp temperatures compared to the actual sample temperature at the beginning of the clamp. The simulation results show that a significant error is only introduced for thick films with large thermal conductivity anisotropy (Fig. S4c). The measurements of single layer films with thicknesses ~1- 8 µm (Fig. S2a) are not significantly affected even for thermal conductivity anisotropy ratios above 200.

### 1.2.2 Time-domain thermoreflectance method

*Sample preparation*

One of the most challenging steps in the time-domain thermoreflectance (TDTR) experiments is sample preparation, whereby a flat and smooth cross-sectional surface of the drawn film has to be created.[35] To this end, we hot pressed (Carver 4120) 100 layers of 50× films into a laminate (~150 µm thick, 1 mm wide and 2 cm long) at 120 °C for 40 minutes. We did not expect dramatic structural change due to the elevated temperature, since the film melting point was measured using differential scanning calorimetry (TA Instruments Discovery) to be ~140 °C (Fig. S5a), consistent with previously reported values (~144 °C) for UHMWPE.[36] We further embedded the laminated 50× film into an epoxy matrix, which was necessary for us to properly mount and cut the cross-section using a microtome (Leica Microsystems). This cutting procedure created a flat and smooth cross-section surface of the laminate and the surrounding epoxy. Using atomic force microscope characterization, the root-man-square roughness of the cross-section surface was measured to be ~10 nm in a 15 μm × 15 μm region (Fig. S6). The difficulty in sample preparation for TDTR is further complicated by the fact that the yield of our drawing platform is not very uniform at higher draw ratio. We thus only prepared one thick sample for the TDTR experiment.

**Heat conduction model**

Measured thermoreflectance signals were fitted to a standard two-dimensional, 3-layer heat conduction model considering the aluminum (Al) transducer, the laminate and in-between interface. Both the out-of-plane (film draw direction) and in-plane thermal conductivity of the UHMWPE laminate were explicitly modeled to account for the expected anisotropy.

**Laser parameters**

A 100-fs-wide pump laser pulse (~400 nm center wavelength) was used to instantly heat up the surface of an aluminum-coated sample (Fig. S6), the cooling of which was then monitored using a probe pulse (800 nm) as a function of delay time between the pulses (Fig. 2c).[26,27] Subsequently, the cooling curves were fitted to a standard two-dimensional heat transfer model to get the sample thermal conductivity (Fig. 2c, d). In order to increase the signal-to-noise ratio, modulated heating was applied by electro-optical modulation of pump power, which resulted in a complex signal with its amplitude and phase recorded by a lock-in amplifier. Both the amplitude and phase signals were used for model fitting. The excellent agreement between amplitude and phase fitting confirmed the measurement reliability (Fig. 2d and Fig. S7). Changing the fitted thermal conductivity by 20% led to a large discrepancy between the simulated and measured curves, further indicating good experimental sensitivity (Fig. 2d and Fig. S8). The reported value in Fig. 3 was obtained as an average of 20 experiments at 3 MHz and 6 MHz modulation.

The incident pump beam power was varied from about 20 mW to 50 mW across multiple measurements, with no pump-power dependence observed. The incident probe power was kept at ~6 mW. At room temperature, such power levels typically result in a small (~1 K) temperature increase so that the change in aluminum reflectance remains proportional to its temperature variation. The $1/e^2$ diameters of the pump and probe beams were measured to be 53 μm and 11 μm, respectively, using a scanning slit beam profiler. Such a configuration ensures that the thermoreflectance signal essentially captures a 1D heat conduction process perpendicular to the sample surface, or equivalently, parallel to the draw direction (Fig. 2c). This setup also helps minimize any uncertainty associated with inaccurate measurement of laser beam size. Finally, multiple pump modulation frequencies (3 MHz and 6 MHz) were used to see if there is any frequency dependence.

**Aluminum transducer layer thickness**

The Al layer thickness was set as 90 nm during electron-beam evaporation and was subsequently measured using a profilometer as 88 ± 3 nm. Thickness was further verified by performing TDTR measurement of a standard sapphire sample, which was Al-coated together with the UHMWPE sample.

**Heat capacity of 50× films**

Differential scanning calorimetry measurement of the heat capacity of the 50× films was performed using a TA Instruments Discovery. The calorimeter was calibrated with a standard sapphire sample prior to measurement of the films. The sample temperature was cycled 3 times between 180 K and 340 K at a heating rate of 5 K/min. The measured specific heat (Fig. S5b) agrees well with literature values,[37] especially near room temperature. Further, a density of 0.97 g/cm$^3$ was used for conversion to volumetric specific heat.[38]

*Measured thermal conductivity and interface conductance*

Thermal conductivity of the laminate (100 layers of 50× films) along the draw direction was measured as 33.6 ± 4.5 W/m-K, in excellent agreement with values obtained using the steady-state method. The Al/UHMWPE interface thermal conductance was simultaneously fitted to be 44.7 ± 4.0 MW/m$^2$-K, which is typical for smooth interfaces at room temperature. In comparison, thermal conductance of a smooth Al/sapphire interface is usually 90 MW/m$^2$-K. The above-mentioned values are averages of 10 individual runs at 3 MHz pump modulation and another 10 at 6 MHz (Fig. S7). When averaged separately, the 3 MHz data yield a film conductivity of 34.8 ± 5.3 W/m-K and an interface conductance of 45.3 ± 5.1 MW/m$^2$-K, while the 6 MHz data yield 32.3 ± 3.1 W/m-K and 44.1 ± 2.5 MW/m$^2$-K, essentially indicating no frequency dependence. We have also verified through sensitivity analysis that both amplitude and phase fitting are reliable for the current study (Fig. S8). Further, the anisotropy ratio of in-plane to out-of-plane thermal conductivity was investigated to have a negligible effect (<1%) on the fitting results.

## 1.2 Structural characterization with synchrotron X-ray

Synchrotron X-ray scattering measurements were carried out at Sector 8-ID-E of Advanced Photon Source (APS), Argonne National Laboratory, with an unfocused and collimated 10.91 keV X-ray beam of cross-section 200×200 μm$^2$ (V×H). Samples were measured in a vacuum chamber in order to minimize radiation damage and scattering background from air and X-ray windows. Single-photon-counting area detector Pilatus 1MF was mounted 137 mm and 2153 mm downstream away from the sample for wide-angle X-ray scattering (WAXS) and small-angle X-ray scattering (SAXS), respectively. Each sample was translated for multiple examination to ensure macroscopic structural homogeneity. Raw WAXS and SAXS patterns were processed with various corrections with MATLAB-based GIXSGUI software before quantitative structural analysis. The film surface was perpendicular to the incident beam as shown Fig. 4a.

### 1.2.1 WAXS and SAXS data corrections

Intensity of the 2D WAXS and SAXS images was corrected on a pixel-by-pixel basis:[39]

$$I_{corrected} = I_{raw} \frac{E_m E_d F C_S}{PL} \quad (7)$$

where $I_{raw}$ is the raw data, $E_m$ is the air gap absorption correction, $E_d$ is the detector efficiency correction, $F$ is the detector's flat-field correction at the operation energy of 10.91 keV, $C_s$ is the solid angle correction, $P$ is the polarization correction, $L$ is the Lorentz correction.

### 1.2.2 WAXS analysis

We identified an orthorhombic cell with lattice constants $a$ = 7.42 Å, $b$ = 4.95 Å and $c$ = 2.54 Å, which agreed well with reference values.[7]

*Effective crystallinity*

X-ray diffraction, or WAXS, is routinely employed to measure the percentage of crystallinity of materials consisting of crystalline and amorphous components. The crystallinity is calculated as the ratio of the integrated intensity from the crystalline peaks to the sum of the crystalline and amorphous intensities.[40]

$$Crystallinity = \frac{I_{crystalline}}{I_{crystalline}+I_{amorphous}} \quad (8)$$

Most samples for crystallinity analysis often take the form of powders. It is quite a challenge to obtain the true value of crystallinity from polymer fiber or sheet samples using this method.[41] This is because while the amorphous components are isotropic, ordered polymer chains and crystalline components often adopt certain orientations, nullifying the isotropy assumption of the method.

In this work, we introduce a concept of "effective crystallinity" which attempts to approximate the true crystallinity. It was calculated with the same Eq. (8) but on a single WAXS image. Since identical WAXS geometry was employed with an area detector for all samples in this work, the trend of the change of the degree of the crystallinity can be qualitatively analyzed as the draw-ratio increases. The limited active area of the detector captures only a little over a quarter of the entire WAXS pattern, i.e. the {$hk0$} and {$hk1$} Bragg groups (Fig. S9a). Considering the symmetry of the scattering which contains four {$hk1$} groups and two {$hk1$} groups, when converting to 1D scattering profile, we are able to extend to the entire WAXS pattern with four quarters by double counting the region containing the {$hk1$} group (Fig. S9a).

The 1D intensity profile was then fitted to the sum of a series of Voigt functions (which is a convolution of Gaussian and Lorentzian functions) to represent each amorphous and crystalline peak, plus a linear background (Fig. S9b, 5× as an example). The integrated intensities of the crystalline peaks were then individually calculated and summed before division by the total intensity to get effective crystallinity. Following such a procedure, an average effective crystallinity of 0.67 was obtained for the 2.5× films, and a crystallinity as high as 0.92 for the 110× films. We note that considerable improvements have been achieved over the years both in the film quality and in our X-ray scattering measurements and analysis.[21]

**Crystallite orientation**

Orientation order parameters: The degree of orientation was quantified by orientation order parameters defined as the intensity-weighted moments of cos$\beta$, where $\beta$ is the tilt angle between the $c$-axis and draw direction (Fig. 4a).[29] The orientation orders can be calculated directly from the measured azimuthal intensity distributions. Perfect alignment of the crystallites results in $\langle \cos^n \beta \rangle = 1$, where $n$ is the order of the moment (Fig. 4d and Fig. S9c).

During stretching, the *c*-axis of the crystallites (Fig. 4a, chain direction) tends to align with the draw direction. Since the *a*, *b*, *c* axes (Fig. 4a) are mutually orthogonal for standard orthorhombic unit cells, and there is no evidence that the *a* and *b* axes prefer any orientations except being orthogonal to the *c*-axis, we thus can determine the orientation order from the azimuthal intensity profiles $I(\phi)$ across the {*hk0*} peaks via an analytical, closed-form method that was developed for liquid crystals.[29] In this method, the 1st, 2nd and 4th order parameters (Fig. S9c) are given by

$$P1 = N^{-1} \int_0^{\pi/2} I(\phi)(\cos^2\phi) d\phi \quad (9)$$

$$\bar{P}2 = 1 - N^{-1} \frac{3}{2} \int_0^{\frac{\pi}{2}} I(\phi)\{\sin^2\phi + (\sin\phi)(\cos^2\phi)\ln[(1+\sin\phi)/\cos\phi]\} d\phi \quad (10)$$

$$\bar{P}4 = 1 - N^{-1} \frac{3}{2} \int_0^{\frac{\pi}{2}} I(\phi) \left\{ \sin^2\phi \left(\frac{105}{16}\cos^2\phi + \frac{15}{24}\right) \right.$$
$$\left. + (\sin\phi)\ln[(1+\sin\phi)/\cos\phi]\left(\frac{105}{16}\cos^4\phi - \frac{15}{4}\cos^2\phi\right) \right\} d\phi$$

(11)

where $N = \int_0^{\frac{\pi}{2}} I(\phi) d\phi$ is the normalization constant (Fig. S9d). Since all {*hk0*} peaks resulted in identical values for the order parameter as we expected, these values were then averaged to give the overall order parameters.

### 1.2.3 SAXS analysis

Small-angle X-ray scattering (SAXS) technique is capable of discovering the structural information of a sample which has spatial inhomogeneities ranging from a few to several hundreds of nanometers. This method is often measured against the scattering wavevector transfer *q*, and the scattering intensity is in principle the modulus square of the Fourier transform of the spatial distribution of the electron density. In kinematic approximation,[30] this relation is $I(\vec{q}) \sim \left| \int d\vec{r} \rho(\vec{r}) e^{-i\vec{q}\cdot\vec{r}} \right|^2$. The contrast of the electron densities between the crystalline and amorphous regions of the polyethylene films, although weak, is sufficient to create enough SAXS signals for the structure analysis. The humps in the SAXS intensity profile along the draw direction signal a periodic structure along that direction; and the *q* positions of the humps scale as 1:2 (Fig. 4e), suggesting a layered or lamellar superlattice packing[30] (alternating crystalline and amorphous phases) with its repeating unit schematically depicted as Fig. S10a.

*Lamellar superlattice structure*

Each repeating unit is modeled as a crystalline layer, an amorphous layer, and two transition layers (Fig. S10a). Since electron density in the transition layers is expected to vary from that of a crystallite to that of the amorphous, it is then reasonable to simplify the model to a two-phase unit of lengths $L'$ and $L-L'$ (Fig. S10a). Rather than assuming a sharp interface as in a three-phase model, the transition between these two phases is effectively modeled as an interfacial roughness taking the form of an error function whose derivative is a Gaussian with a standard deviation denoted by σ. This model creates a smooth transition between crystal and amorphous layers, which is more realistic than the two-step functions in the three-phase model (Fig. S10).

**Electron density profile**

The electron density profile in each repeating unit along the *c*-axis (Fig. S10b) is thus a smooth profile $\rho(x)$ determined by $L$, $L'$ and σ as $\rho(x) = \frac{\sum_{j=1}^{3} \rho_j W_j(x)}{\sum_{j=1}^{3} W_j(x)}$, where $\rho_1 = \rho_3 = \rho_A$ and $\rho_2 = \rho_C$ are the electron densities of the amorphous and crystalline phases, respectively, and

$$W_1(x) = \frac{1}{2}\left[1 + \text{erf}\left(\frac{x-\frac{L'}{2}}{\sqrt{2}\sigma}\right)\right] \quad (12)$$

$$W_2(x) = \begin{cases} \frac{1}{2}\left[1 + \text{erf}\left(\frac{x}{\sqrt{2}\sigma}\right)\right], & x \leq 0 \\ \frac{1}{2}\left[1 - \text{erf}\left(\frac{x}{\sqrt{2}\sigma}\right)\right], & x > 0 \end{cases} \quad (13)$$

$$W_3(x) = \frac{1}{2}\left[1 - \text{erf}\left(\frac{x+\frac{L'}{2}}{\sqrt{2}\sigma}\right)\right] \quad (14)$$

The Fourier transform of $\rho_L(x)$ is called the form factor in SAXS

$$F(q) = \int dx \rho(x) e^{-iqx} \quad (15)$$

A Gaussian distribution with a standard deviation of $\sigma_{L'}$ for the length $L'$ is introduced to account for its polydispersity (Fig. S10b),

$$g(L'_\alpha) = \frac{1}{\sqrt{2\pi\sigma_{L'}^2}} \exp\left[-\frac{(L'_\alpha - L')^2}{2\sigma_{L'}^2}\right] \quad (16)$$

**Structure factor and size distribution**

The structure factor $S(q)$ describing how the units are stacked to form a lamellar superlattice, is modeled within the framework of the 1D para-crystal model,[42] so that the long-range ordering is gradually destroyed in a probabilistic manner (which is often modeled as a Gaussian).[43] This model allows a link between perfectly ordered and disordered structures, and provides a good analytical description of a structure in a partially ordered state. In this 1D para-crystal model,[44] the distance between successive units is independent of other distances and obeys a statistical distribution function $p(x)$ with $\int_{-\infty}^{\infty} dx p(x) = 1$. Assuming a Gaussian probability density, we have

$$p(x) = \frac{1}{\sqrt{2\pi\sigma_D^2}} \exp\left[-\frac{(x-D)^2}{2\sigma_D^2}\right] \quad (17)$$

whose Fourier transform is

$$P(q) = \exp(-q^2\sigma_D^2/2)\exp(iqD) \quad (18)$$

The autocorrelation function $g(x)$ of the positions of the superlattice units is given by

$$g(x) = \delta(x) + g^+(x) + g^-(x) \quad (19)$$

where $\delta(x)$ is the Dirac delta function, $g^+(x)$ and $g^-(x)$ are the autocorrelation functions on the positive ($x > 0$) and negative ($x < 0$) sides of the axis, respectively, which are written as

$$g^+(x) = p(x) + p(x) \otimes p(x) + p(x) \otimes p(x) \otimes p(x) + \cdots \text{ for } x > 0 \quad (20)$$

and by symmetry

$$g^-(x) = g^+(-x) \text{ for } x < 0 \quad (21)$$

Here $\otimes$ denotes the convolution. The structure factor is defined as the Fourier transform of $g(x)$ and thus given by

$$S(q) = 1 + P(q) + P^2(q) + P^3(q) + \cdots = \text{Real}\left[\frac{1+P(q)}{1-P(q)}\right] \quad (22)$$

For lamellar superlattice stacking, $D = L$, and the distance of successive units along the drawing direction is coupled with the unit lengths, rather than being independent as in the pure 1D para-crystal model. We therefore adopt the size-spacing coupling approximation (SSCA),[44,45] which includes a size-spacing coupling parameter $\kappa$ with $\kappa = 0$ automatically reduces to the decoupling approximation given in the classical para-crystal model. The structure factor in SSCA is

$$S(q) = 1 + 2\text{Real}\left\{\frac{\tilde{P}_\kappa^2(q)\Omega_\kappa(q)}{\tilde{P}_{2\kappa}(q)[1-\Omega_\kappa(q)]}\right\} \quad (23)$$

with the scattering intensity at $q \neq 0$ given by

$$I(q) \sim |\langle F(q)\rangle|^2 + 2\text{Real}\left[\tilde{F}_\kappa(q)\tilde{F}_\kappa^*(q)\frac{\Omega_\kappa(q)}{\tilde{P}_{2\kappa}(q)[1-\Omega_\kappa(q)]}\right] \quad (24)$$

where

$$\tilde{P}_\kappa(q) = \int dL'_\alpha g(L'_\alpha)\exp[i\kappa q(L'_\alpha - L')] \quad (25)$$

$$\tilde{F}_\kappa(q) = \int dL'_\alpha\, g(L'_\alpha)F(q, L'_\alpha)\exp[i\kappa q(L'_\alpha - L')] \quad (26)$$

$$\Omega_\kappa(q) = \tilde{P}_{2\kappa}(q)P(q) \quad (27)$$

and $\langle F(q)\rangle$ is the polydispersity averaging

$$\langle F(q)\rangle = \int dL'_\alpha g(L'_\alpha)F(q, L'_\alpha) \quad (28)$$

Fig. 4e shows the best fit of the model to SAXS data, with its continuous electron density contrast profile shown in the inset of Fig 4f. Figure S11a shows the structure factor and Fig. S11b shows the $L'$-size distribution along the fiber direction with different draw ratio ($L'$ is the length of crystal and transition region, Fig. S10) As draw ratio increases, the humps move to small $q$, i.e. larger length scales, and also become less significant (i.e. more structural disordering as shown in Fig. S11b for high draw ratios).

The fiber diameter can be statistically estimated by analyzing the SAXS intensity along the meridian direction. This intensity is given by two contributions as the new Guinier-Porod model described[46],

$$I(q) = \frac{G}{q^s} \exp\left(\frac{-q^2 R_g^2}{3-s}\right), \text{ for } q \leq q_1,$$
$$I(q) = \frac{D}{q^d}, \text{ for } q \geq q_1,$$

where

$$q_1 = \frac{1}{R_g}\left[\frac{(d-s)(3-s)}{2}\right]^{1/2},$$
$$D = G\exp\left(\frac{-q_1^2 R_g^2}{3-s}\right) q_1^{(d-s)}.$$

Here, G and D are Guinier and Porod scale factors, respectively, d is the Porod exponent, s is the dimensionality parameter, $q_1$ is the boundary wave vector transfer where Guinier and Porod contributions merge, and $R_g$ is the radius of gyration. In this case, $R_g$ is cross-sectional radius of gyration of the cylinders if we approximate the aligned fibers as a bundle of cylinders. The mean radius of the fibers is thus given by $R = \sqrt{2} R_g$. Figure S11c shows the fitting data by new Guinier-Porod model, and diameter $R_g$ =7.8 nm, R = 11.1 nm are obtained for 70× sample.

### 1.2.4 All the WAXS and SAXS patterns

Figure. S12 and Fig. S13 show all the WAXS and SAXS patterns for the as-extruded (1×) and drawn films with different draw ratios, respectively.

### 1.3 Thermal conductivity model

We employed a one-dimensional heat transfer model to compute the film thermal conductivity, which depends on the thermal conductivities of the crystalline and amorphous regions, as well as the amorphous fraction ($\eta$) in one periodic unit (amorphous length / period length, Fig. S10). This model was built to unambiguously pinpoint the key role of the amorphous region, specifically how the thermal conductivity in the amorphous part changes at high draw ratios. The approximation that the thermal transport can be described by a 1D model is reasonable for high draw ratios, due to the fact that SEM observations and SAXS results (Fig. 1g-j, Fig S3c, and Fig. S11c) suggest a very small fiber diameter ~8 nm. Such a 1D model contains a unit cell of total length $L$ including a crystalline and an amorphous region. We note that the transition region as discussed before has been effectively included into the crystalline region (Fig. S14 and Fig. S10a). Adding all the heat resistances, the effective thermal conductivity of this 1D model is then given by $k = [(1-\eta)/k_c + \eta/k_a]^{-1}$, as already shown in the main text.

We first discuss the extraction of the structural parameters used in the equation. As mentioned above, the SAXS measurement characterizes periodicity of the lamellar superlattice. One can first obtain the period length from SAXS structure factor analysis. The lengths (and ratios) of different regions (amorphous / crystalline) can be further estimated by studying the electron density distribution. In Fig. 4f inset, we show the electron density profile with respect to the draw ratio measured on different samples. The middle region becomes larger and larger as draw ratio increases, and can be identified as the crystalline region, while the two sides where the electron density is close to zero are the amorphous regions. We take the amorphous region as the part where the electron density is less than 0.05 (corresponding to a length $L_A$). The remaining part ($L$-$L_A$) is taken as the effective crystallite size $L_C$. The ratio between the amorphous region and the total length $L_A/L$ gives the amorphous fraction $\eta$ (Fig. 4f, circles).

Due to the uncertainties involved in estimating the lengths from the SAXS measurement data, we do not directly use the experimental data points for $\eta$ and $L_C$ in the thermal model. Instead, we fit these values by a simple functional form - $C_1*n^{C_2}$ ($n$ being the draw ratio), which should generally capture the trend of the structural parameters as the draw ratio. For the amorphous fraction, the fitting to the experimental values yields $C_1 = 0.6$ and $C_2 = -0.43$. To take into account the large uncertainty associated with these values, 40% variations have been added to them, which translate into the upper and lower bounds for the estimated $\eta$ as shown in Fig. 4f (shaded region). Similarly, for the crystallite size, we obtained $C_1 = 7\ nm$ and $C_2 = 0.28$, with a 20% variation considered that leads to a range for the estimated crystallite size as the draw ratio, shown in Fig. S15a.

Thermal conductivity of the crystalline region was taken from the literature, where the size-dependent thermal conductivity of a single polyethylene chain was calculated using the first principles method.[18] This is then combined with our fitted crystallite size (Fig.S15b) to yield an estimation for the crystalline thermal conductivity at different draw ratios, as shown in Fig. S15b. Here we employed the single chain thermal conductivity as our estimation for the thermal conductivity in the crystalline part because it presents the higher limit. We note that, this together with the fact that the transition region is assumed to have the crystalline thermal conductivity, overestimates the thermal conductivity in the crystalline and transition regions. Therefore, our extracted thermal conductivity for the amorphous region will only be underestimated.

We finally note that, in plotting Fig. 3b, the shaded region is obtained by fitting the measured total thermal conductivity with a straight line ($3.8 + 0.5 \times n$, W/m-K). The upper and lower bounds originate from the uncertainties we have considered in the estimation of crystallite sizes as well as the amorphous fractions, as mentioned above.

# 2. Supplementary table

*Table S1. Measured film dimensions* at various draw ratios.

| Draw ratio | Length (mm) | Cross-sectional area (mm$^2$) |
|:---:|:---:|:---:|
| 1 | 1.3 | 0.6876 |
| 2.5 | 1 | 0.10395 |
| 5 | 1.1 | 0.06435 |
| 10 | 2 | 0.055858 |
| 20 | 1.5 | 0.050818 |
| 30 | 2 | 0.028584 |
| 40 | 2.2 | 0.043407 |
| 50 | 2.25 | 0.021630 |
| 60 | 2.2 | 0.016458 |
| 70 | 1.65 | 0.0083974 |
| 80 | 1.15 | 0.010078 |
| 90 | 1.1 | 0.0075624 |
| 100 | 0.65 | 0.0068800 |
| 110 | 1 | 0.0052500 |

# 3. Supplementary figures

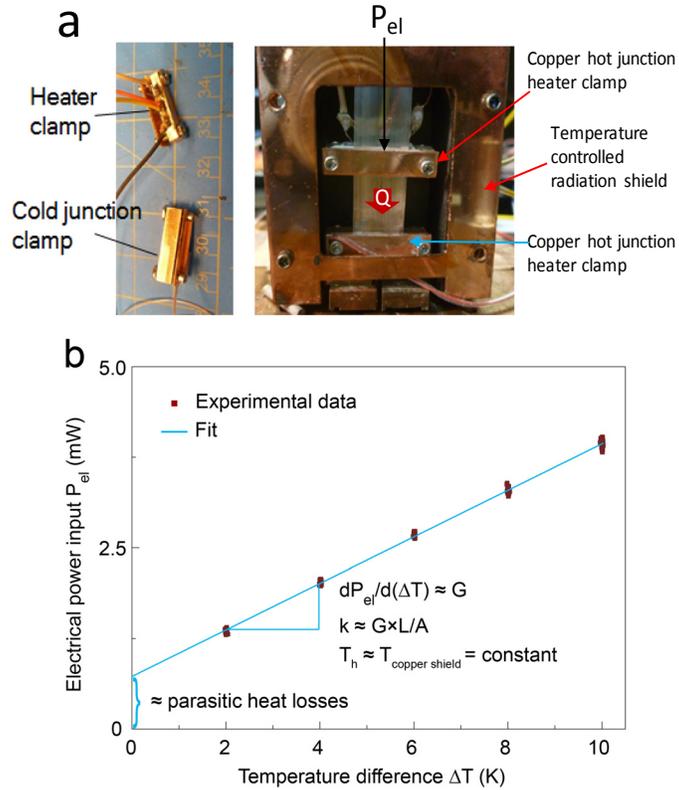

**Fig. S1. Differential steady-state method to measure thermal conductivity.** (**a**) Experimental setup photos showing a plastic film suspended between the hot and cold copper clamps, as well as the temperature-controlled copper radiation shield. (**b**) Representative data from measurement of a 50× film illustrating the differential nature of the method. Electrical heater power is measured at a series of temperature differences. The temperature difference is varied by changing the cold side temperature while maintaining the heater temperature constant. The linear fit slope corresponds to the thermal conductance, $G = kA/L$, of the sample.

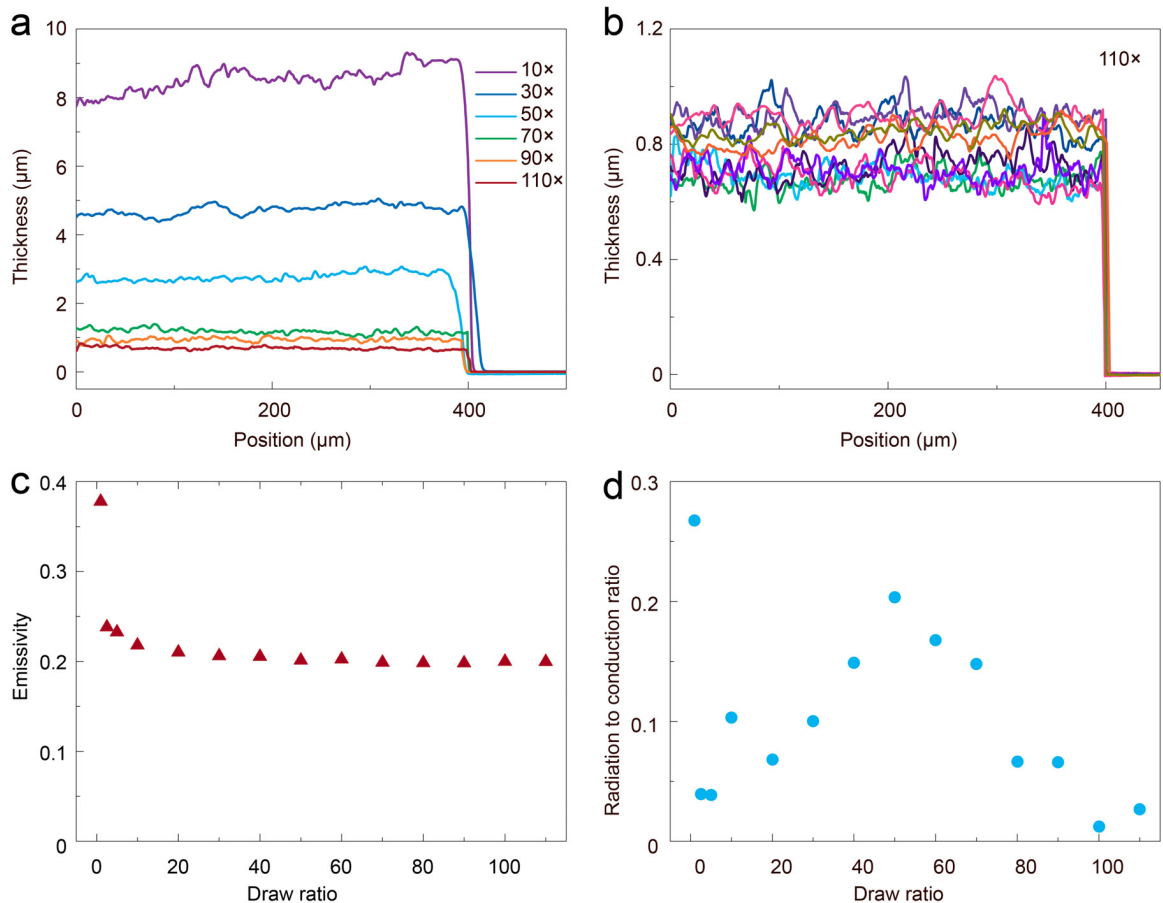

**Fig. S2. Polyethylene film thickness and radiation error analysis.** (**a**) Drawn polyethylene film thickness is in the range of ~ 1-8 $\mu m$, as measured by a stylus profilometer. (**b**) Representative profiles for 110× film at 10 different locations along the sample. (**c**) Computed emittance for various polyethylene films at 298 K. (**d**) Radiation error as a function of draw ratio. In general, radiation errors are less than 20%, and for the higher draw ratios, less than 10%.

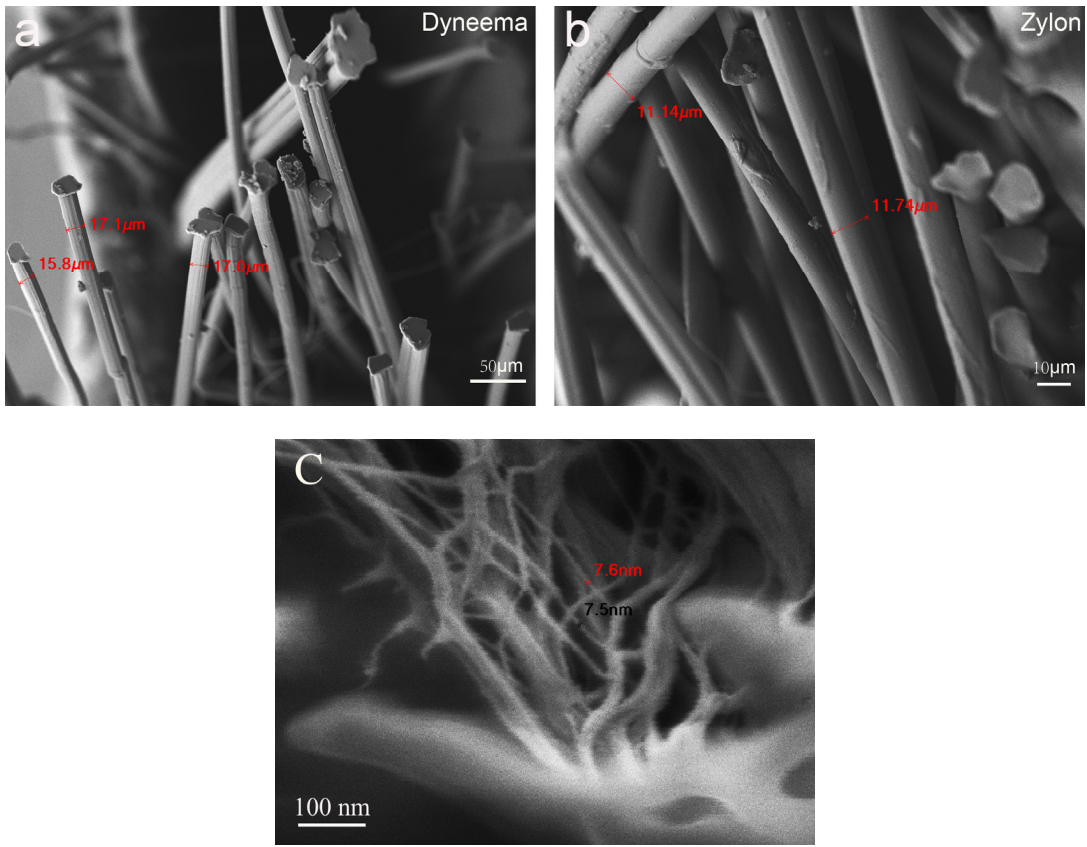

**Fig. S3. SEM images.** (**a**) Average Dyneema fiber diameter is ~17 $\mu m$. (**b**) Zylon fiber diameter is ~11.7 $\mu m$. (**c**) Images of a torn 70× polyethylene film in this work, the interior nanofiber diameters is ~8 nanometers.

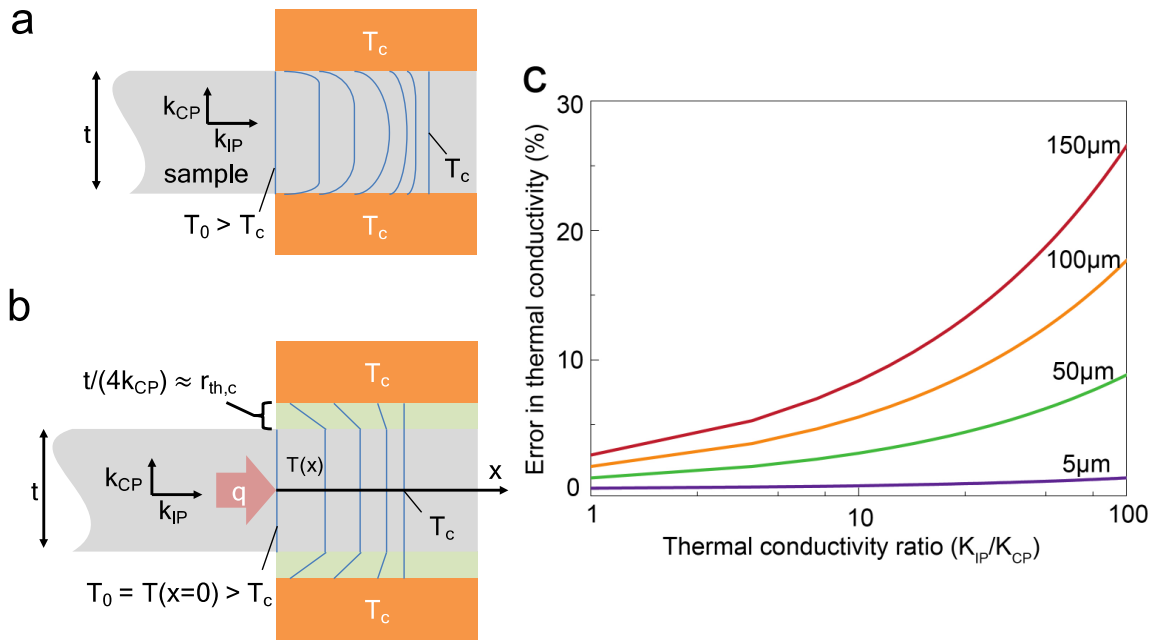

**Fig. S4. Effect of anisotropic thermal conductivity of drawn polyethylene films.** (**a**) Temperature profile along the sample thickness is non-uniform in the clamped region, and temperature on the left side of the clamp is inevitably higher than that of the clamp. (**b**) Simplified model to estimate effect of the cross-plane (2D) fin resistance on the thermal conductivity results. Cross-plane thermal resistance of the sample is lumped into an effective thermal contact resistance $r_{th,c} \approx t/4k_{cp}$. (**c**) Modeling results for a 4-mm long sample show a small error for thin film thicknesses (< 5 µm), even when the thermal conductivity anisotropy ratio is more than 200. The error associated with anisotropy dramatically increases for thicker films. Therefore, to minimize the fin resistance effect on the thermal conductivity measurement the sample should be long and thin.

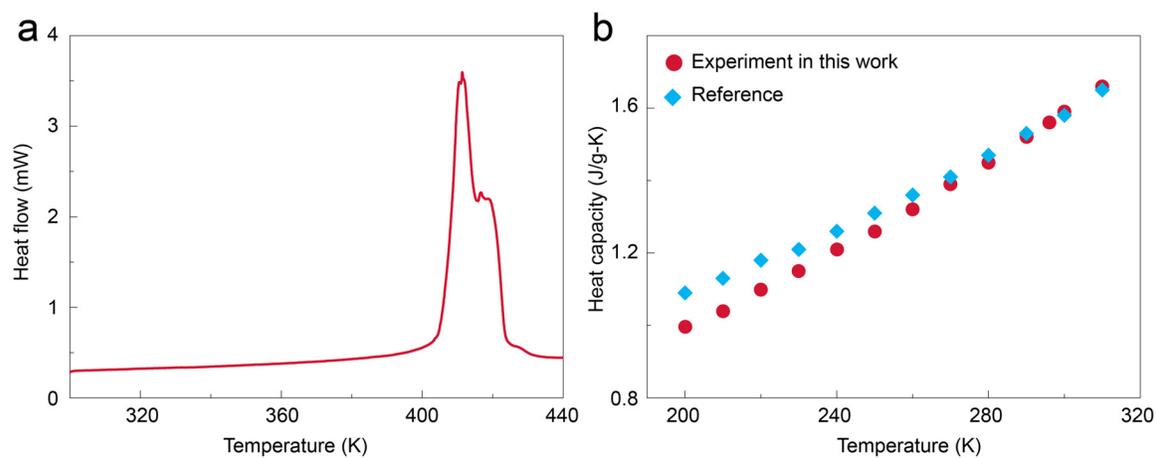

**Fig. S5. Melting temperature and specific heat capacity.** (**a**) Melting temperature of the 50× films is ~140 °C, as indicated by the peak in heat flow using differential scanning calorimetry. (**b**) Specific heat capacity of 50× films as a function of temperature. Measured specific heat of the 50× films agree well with previously reported values,[37] especially near room temperature. The average data from three temperature cycles are plotted.

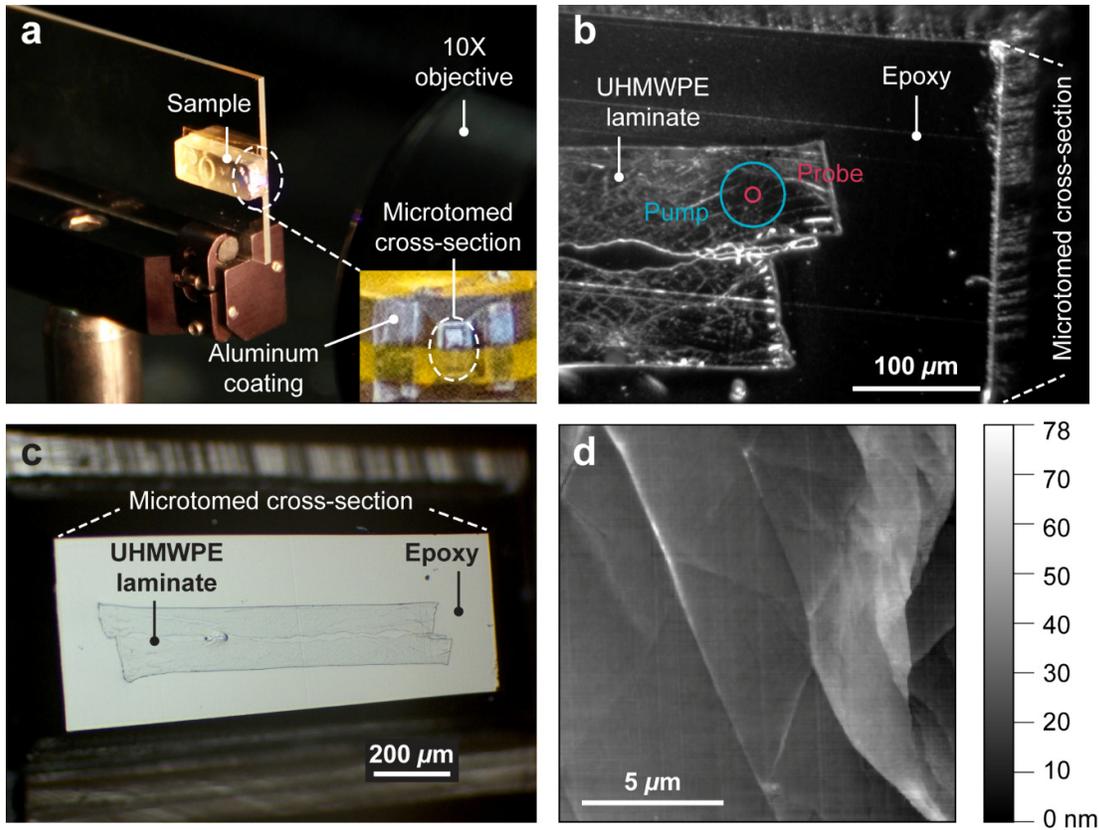

**Fig. S6. Images across multiple length scales of a TDTR sample.** The sample features a UHMWPE laminate embedded in an epoxy matrix, and was carefully cut with a microtome at room temperature in order to reveal a flat and smooth cross-section for TDTR measurement. (**a**) Photo of the sample mounted in front of a long-working-distance 10× microscope objective. Inset is a zoom-in view of the microtomed cross-section partially coated with an 88 nm-thick aluminum layer. (**b**) Dark-field optical micrograph of the sample cross-section obtained during a TDTR measurement. The UHMWPE laminate cross-section is ~1 mm × 150 $\mu$m and consists of 100 layers of as-drawn 50× films hot pressed together. It separated into two halves during sample preparation. Smooth and dark regions generally indicate good sample surface quality. The blue and red circles show the pump (53 $\mu$m in diameter) and probe (11 $\mu$m) spots, respectively. (**c**) Bright-field optical micrograph of the cross-section prior to e-beam evaporation of aluminum. (**d**) AFM image of the UHMWPE laminate cross-section. The root-mean-square surface roughness is ~10 nm.

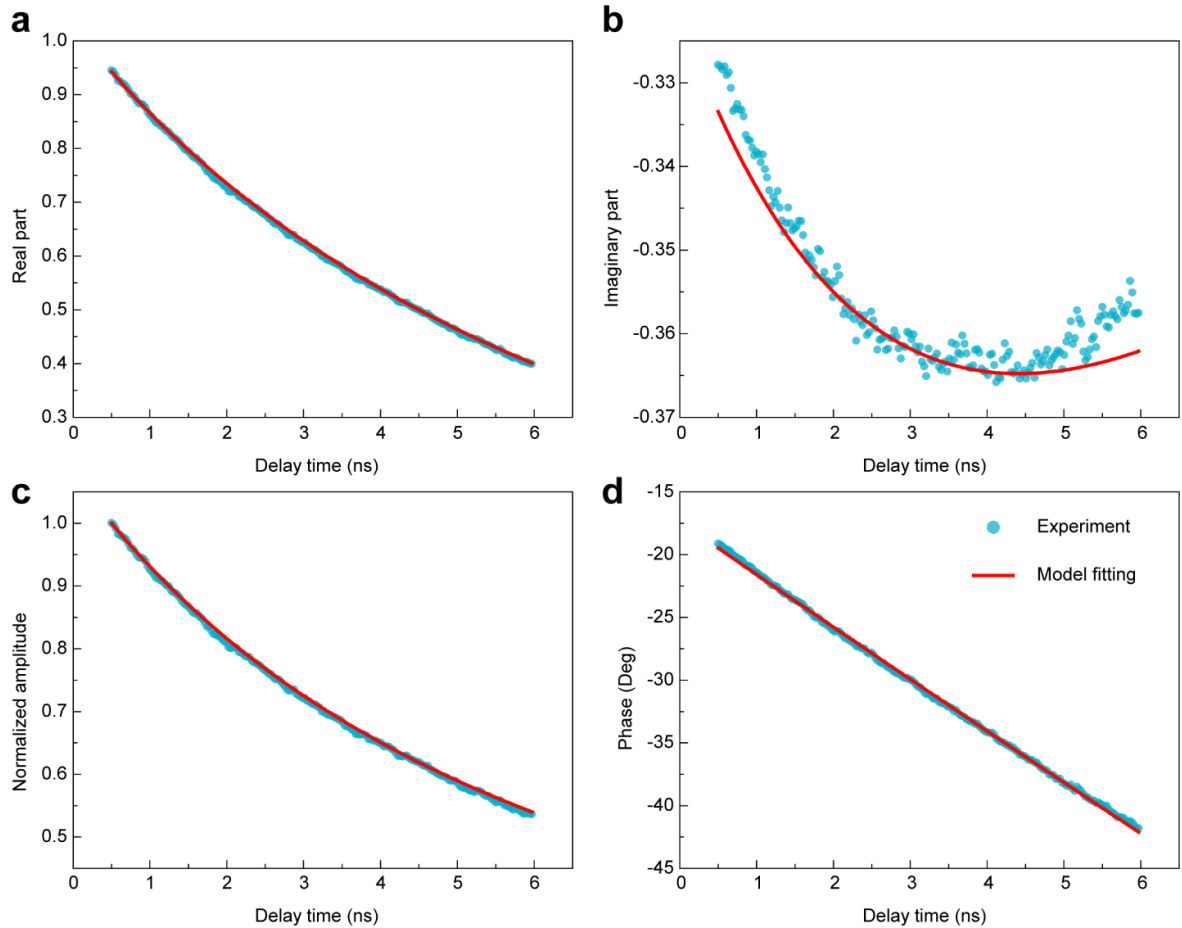

**Fig. S7. Measured and fitted complex thermoreflectance signals.** The measured data is the average of 10 individual runs using a modulation frequency of 6 MHz. (**a-b**) Real and imaginary parts of the complex signal, respectively. (**c-d**) Amplitude (normalized) and phase representation of the same signal, respectively. Phase fitting was performed to obtain the sample thermal conductivity together with the aluminum/sample interface thermal conductance, which were subsequently used to compute all the red curves. The reliability of the experimental results is demonstrated by the fact that fitting to phase alone leads to excellent agreement between modeled and measured data in all four panels. As expected, fitting of the amplitude yields equally good results (see Fig. 2d).

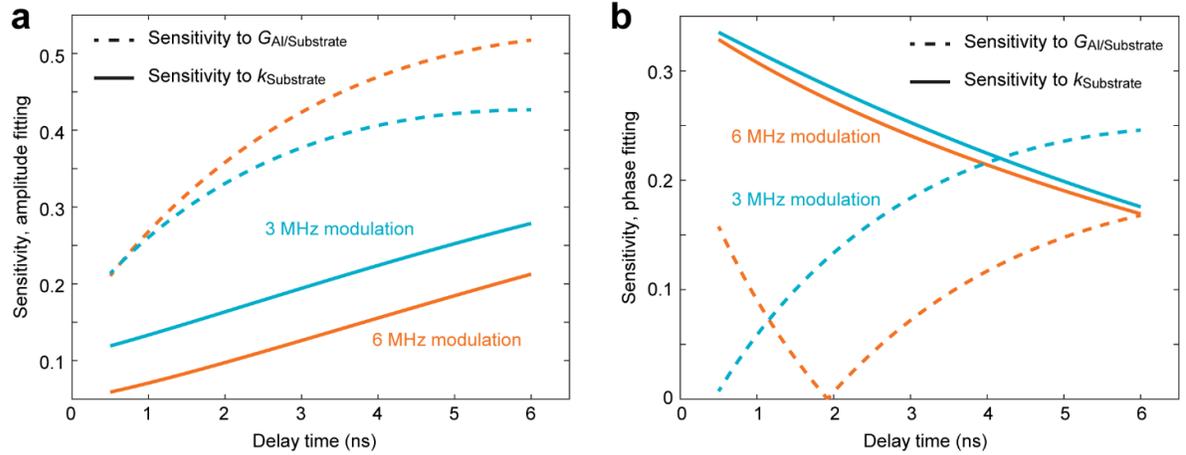

**Fig. S8. Sensitivity analysis of the TDTR experiment.** (**a**) Sensitivity[26,27] of the amplitude of the complex thermoreflectance signal to the aluminum/substrate (UHMWPE laminate) interface thermal conductance ($G_{Al/Substrate}$) and substrate thermal conductivity ($k_{Substrate}$). (**b**) Sensitivity when fitting to the phase of the thermoreflectance signal. Although amplitude fitting and phase fitting offer different relative sensitivity to the substrate conductivity and the interface conductance, both are sufficiently sensitive considering the relatively small experimental noise and more importantly the excellent agreement between results from phase and amplitude fitting.

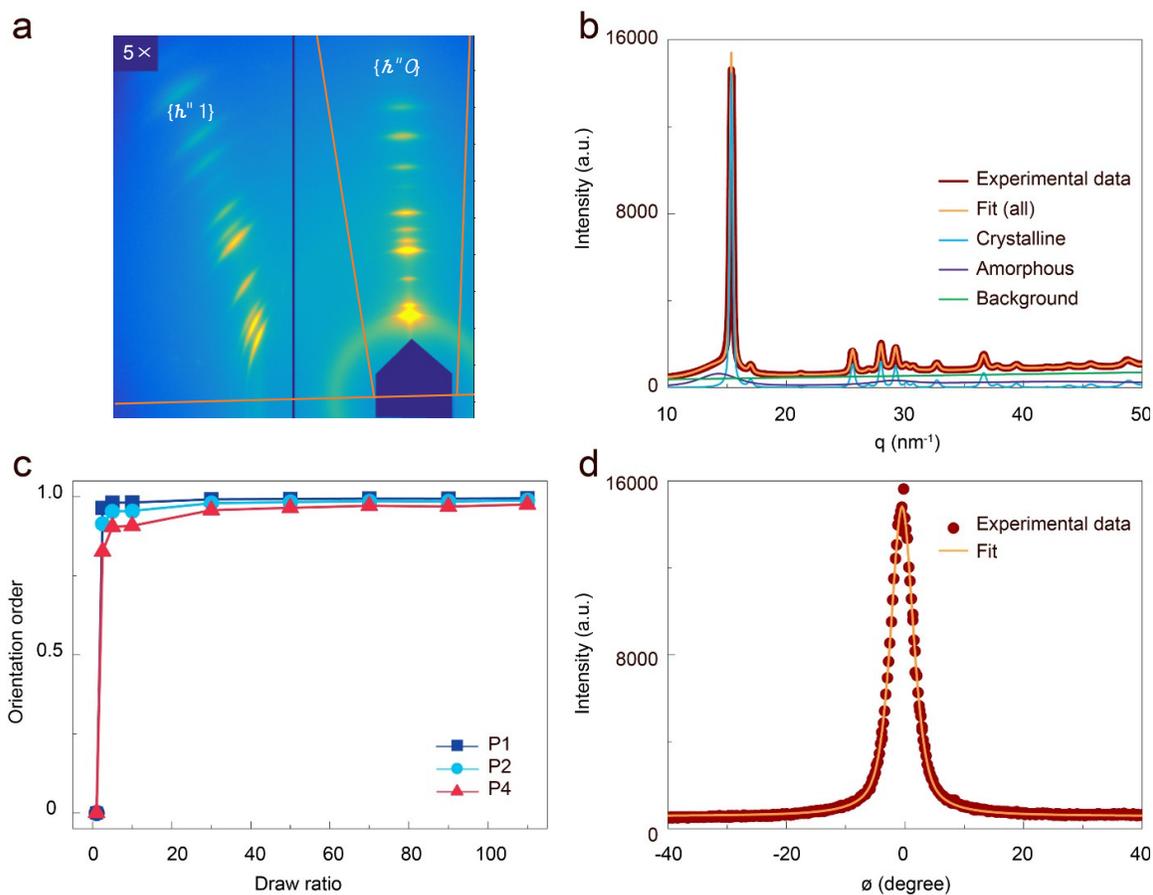

**Fig. S9. Effective crystallinity and crystallite orientation analysis from WAXS.**
(**a**) WAXS pattern of the 5× films, the limited active area of the detector captures only a little over a quarter of the entire WAXS pattern, i.e. the {*hk0*} and {*hk1*} Bragg groups. (**b**) 1D profile with best fits of the crystalline, amorphous and a linear background peaks by Voigt model. (**c**) Orientation order parameters of the films at different draw ratios. (**d**) The intensity profile from the (*200*) peak as a function of the azimuthal angle ∅. The peak is first fitted to a Voigt function, which is then used for the calculation of the order parameters.

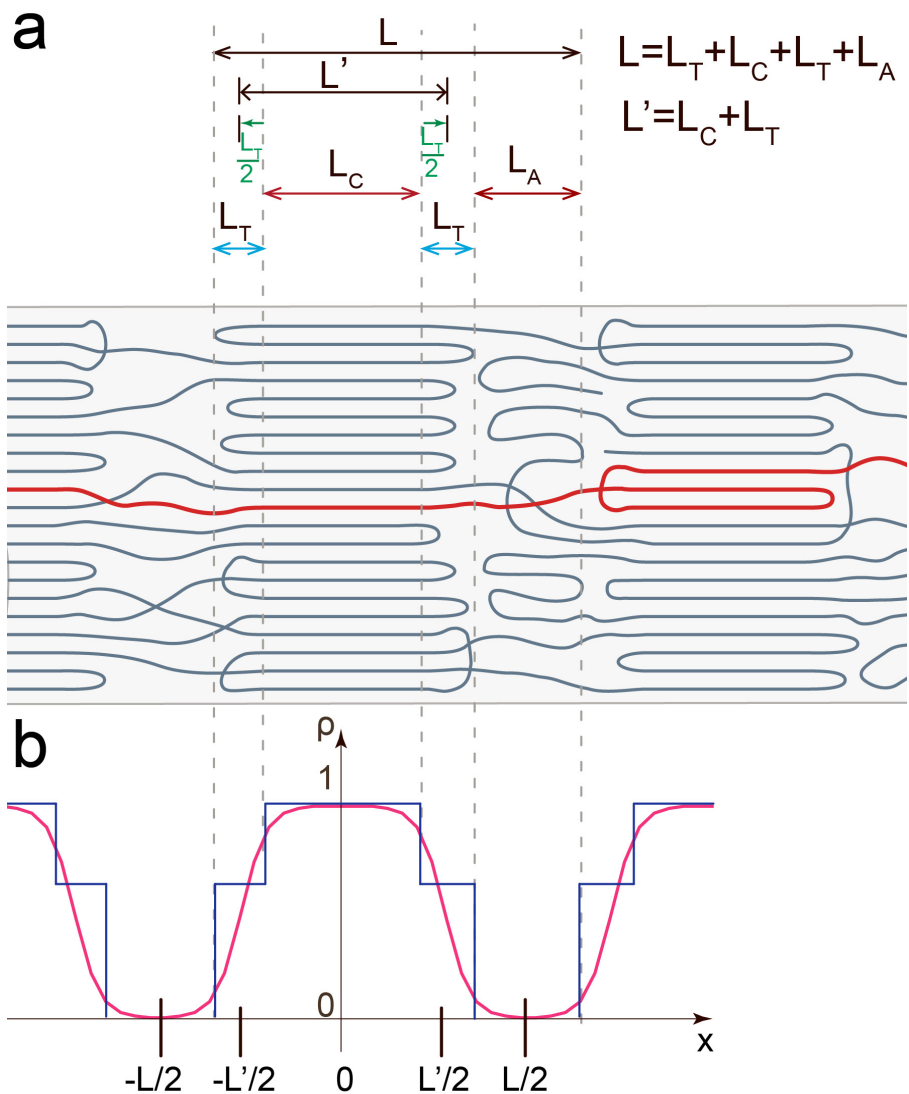

**Fig. S10. Schematic of the lamellar superlattice structure and electron density profile.** (**a**) Each repeating unit of the lamellar superlattice is of a total length $L$, which includes three phases: a crystal phase of length $L_C$, an amorphous phase of length $L_A$, and two transition layers of length $L_T$ in between to represent an electron density change from crystalline to amorphous. (**b**) Step-like three phase model describing the electron density profile of the unit is simplified by a two-phase model with a continuous density profile. In the two-phase model, the transition is modeled by an error function profile (whose derivative is a Gaussian of standard deviation of $c$).

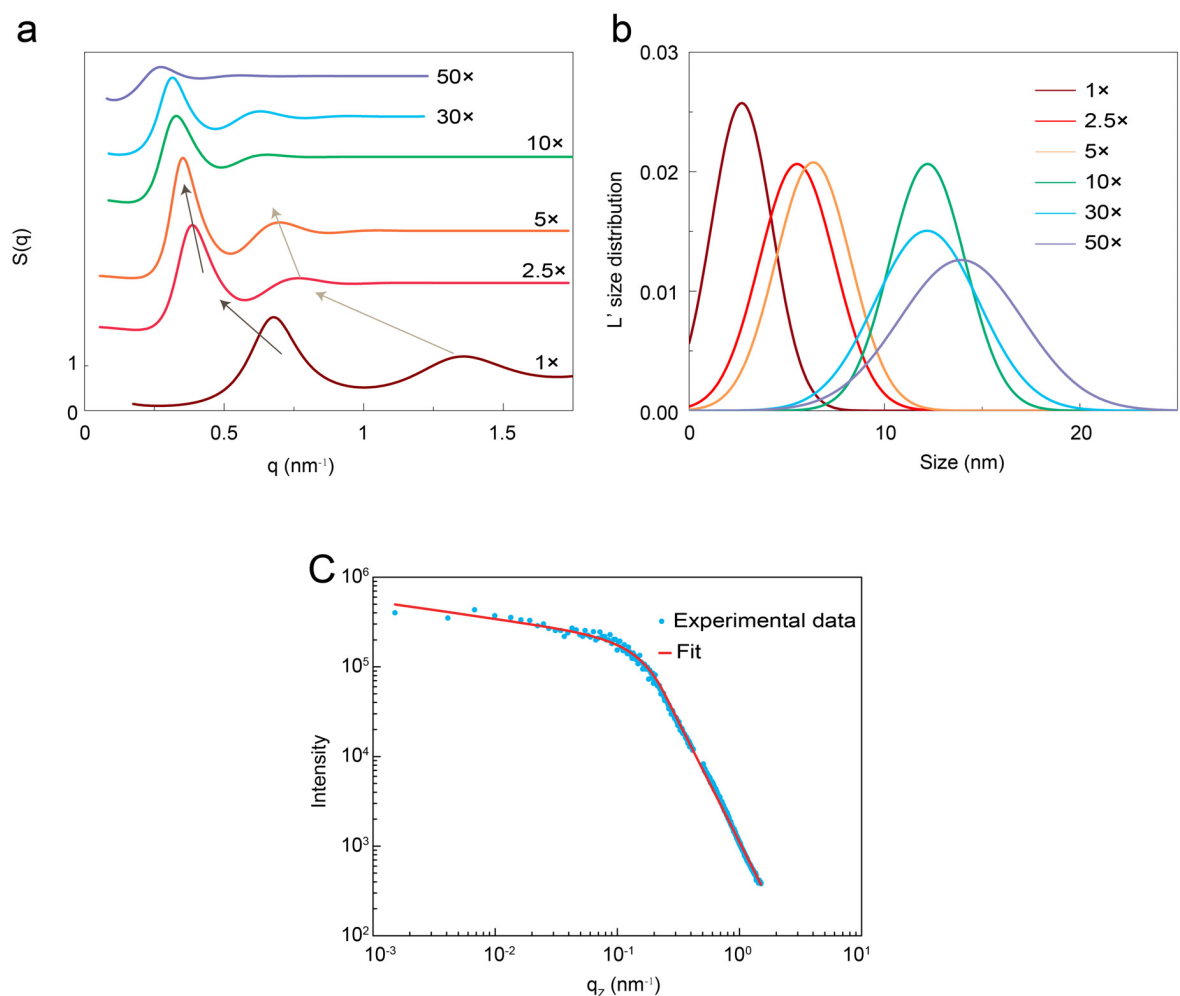

**Fig. S11. The structure factor and size distribution analyzed by SAXS.** (**a**) The structure factor $S(q)$ is modeled using the 1D para-crystal model, and describes how the units are stacked to form a lamellar superlattice. The curves are vertically shifted for clarity (except 1×). As draw ratio increases, the humps move to smaller $q$, suggesting larger length scales in the higher draw ratio films. The humps become less significant at higher draw ratios, suggesting more structural disordering as shown in Fig. S11b. (**b**) The length distribution of L' phases along the fiber direction at different draw ratios (L' is the length of crystal and transition region shown in Fig. S10). (**c**) Statistical analysis of fiber diameter, the fitting curve for 70× sample by Guinier-Porod model. The obtained diameter of nanofiber is ~11.1 nm.

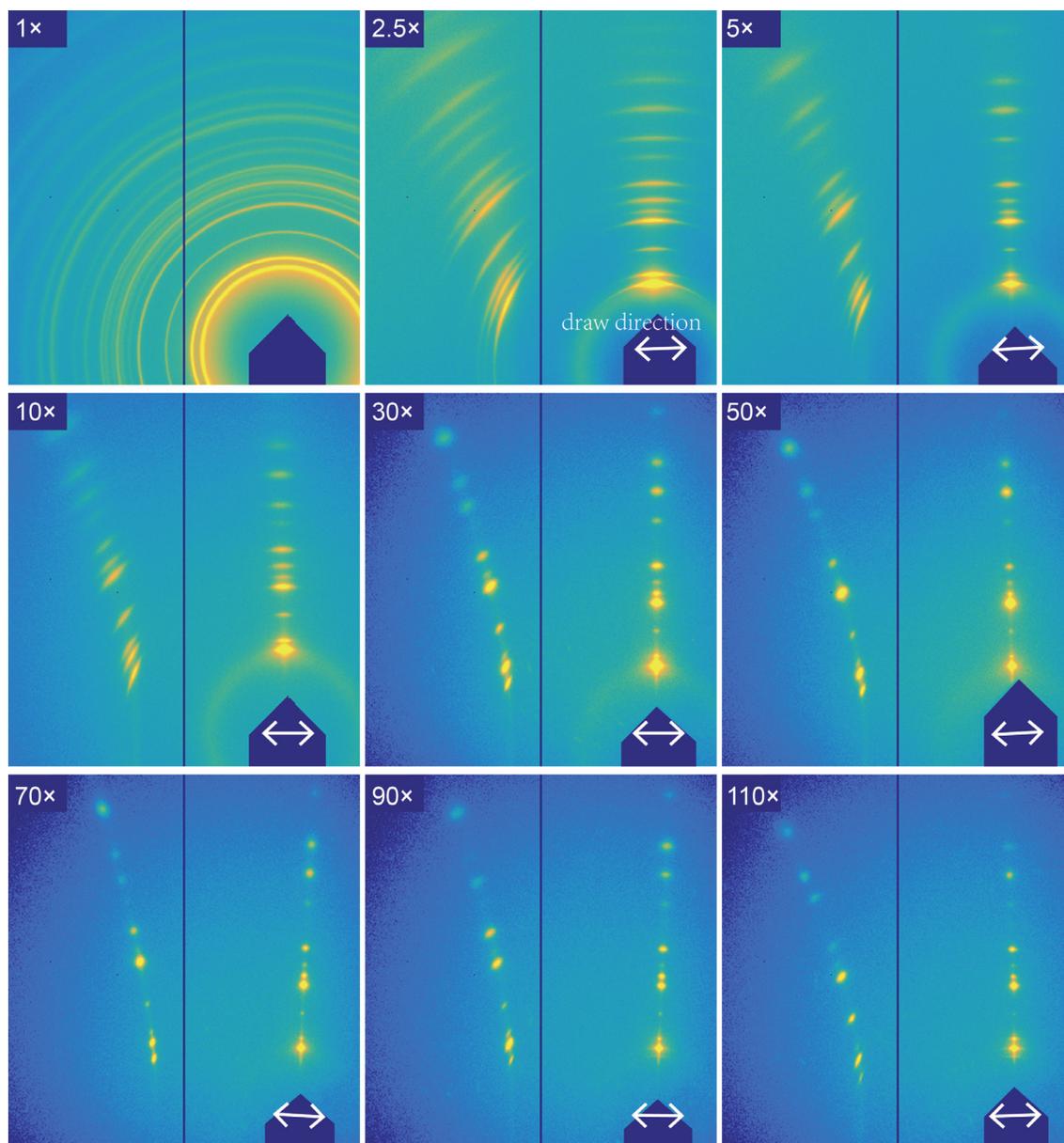

**Fig. S12. WAXS patterns collected for 1× to 110× polyethylene films.** The as-extruded (1×) film is isotropic, while the drawn films (from 2.5× to 110×) all show some anisotropic features. Detailed analysis can be found in supplementary method.

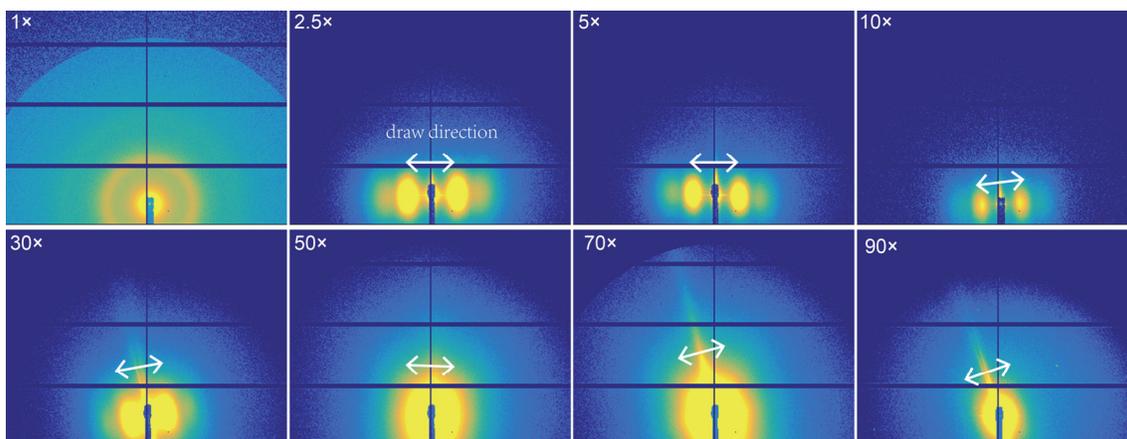

**Fig. S13. SAXS patterns collected for 1× to 90× polyethylene films.** The as-extruded (1×) film is isotropic, while the drawn films (from 2.5× to 90×) all show some anisotropic features. Detailed analysis can be found in supplementary method.

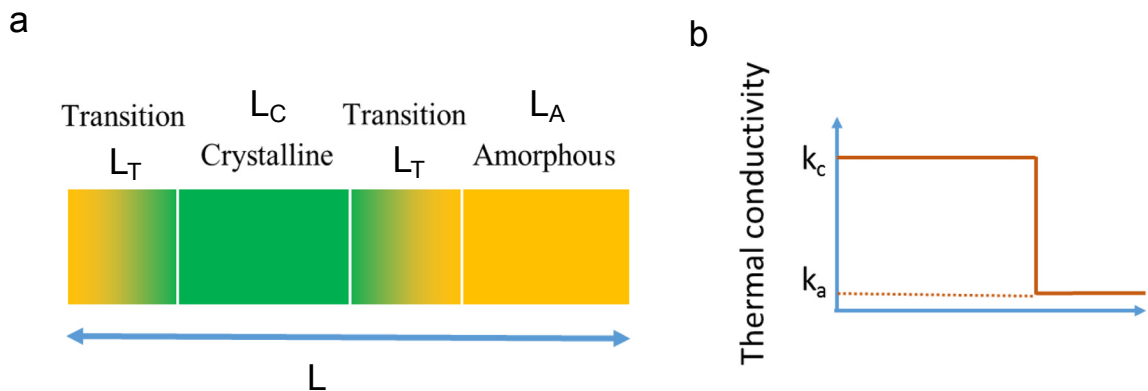

**Fig. S14. Thermal conductivity model.** (**a**) Schematic of the 1D model, a simplification of Fig. S10; (**b**) Modeled variation of thermal conductivity along chain direction in one unit cell, noting that the transition region has been included into the crystalline part by assuming that the former has a thermal conductivity equal to that of the latter.

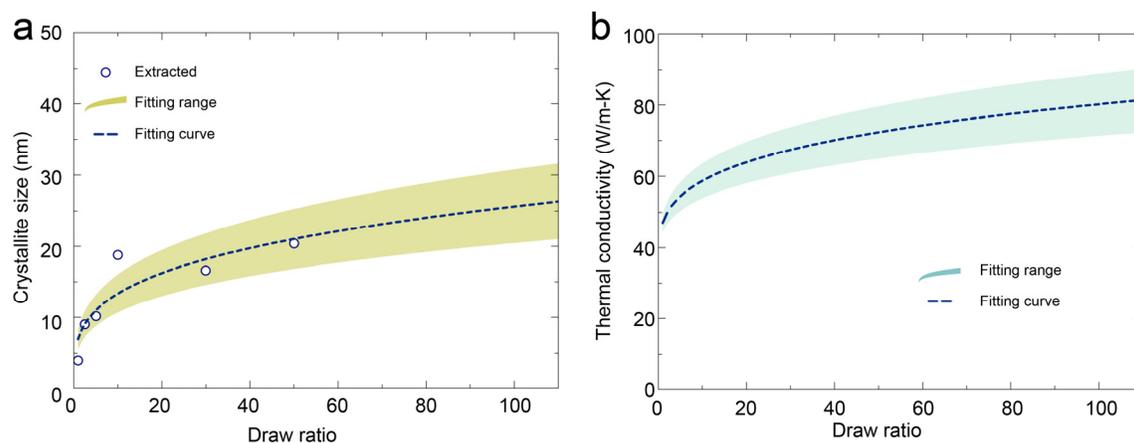

**Fig. S15. Crystallite size and crystalline thermal conductivity model.** (a) Crystallite size at different draw ratios; (b) crystalline thermal conductivity at different draw ratios. The shaded regions are the fitted range, and the dashed line represents the fitting curve.